%% file: fscd_arxiv.tex
\documentclass[11pt, letterpaper]{article}
\usepackage[margin= 1in]{geometry}
\usepackage{authblk}

\usepackage{titlesec}
\usepackage{setspace}

\usepackage[natbibapa]{apacite}
\usepackage{doi}

\usepackage{mathtools}
\usepackage{amssymb}
\usepackage{amsmath}

\usepackage{graphicx}
\usepackage[singlelinecheck= false]{caption}
\usepackage{caption}
\usepackage{subcaption}
\usepackage{rotating}
\usepackage{booktabs}
\usepackage{threeparttable}
\usepackage{hyperref}

\DeclareMathOperator{\N}{\mathcal N}
\DeclareMathOperator{\Be}{Beta}
\DeclareMathOperator{\Dir}{Dirichlet}
\DeclareMathOperator{\Gam}{Gamma}

\newcommand*{\dbar}[1]{\bar{\bar{#1}}}
\newcommand*{\vbar}{\,\vert\,}

\newcommand*{\E}{\mathrm{E}}

\input{notation-fscd.tex}

\hypersetup{colorlinks= true, allcolors= [rgb]{0.0,0.29,0.60}}

\titleformat{\subsection}[runin]{\bfseries}{\thesubsection.}{0.5em}{}[.]
\titleformat{\subsubsection}[runin]{\bfseries\boldmath}{\thesubsubsection.}{0.5em}{}[.]

\title{A Flexible Stochastic Conditional Duration Model\footnote{We thank David Benatia, the audiences of the European Seminar on Bayesian Econometrics (ESOBE) conference in New Orleans, the Seminar on Bayesian Inference in Econometrics and Statistics (SBIES) conference in Providence, and the Computing in Economics and Finance conference in Ottawa for their valuable comments. Email addresses: samuel.gingras@umontreal.ca (S.\ Gingras), william.j.mccausland@umontreal.ca (W.J.\ McCausland)}}

\author[1]{Samuel Gingras\thanks{Corresponding author: D\'epartement de sciences \'economiques, Universit\'e de Montr\'eal, C.P.\ 6128, succursale Centre-Ville, Montr\'eal, QC H3C 3J7, Canada. Phone: +1 514-343-6539}}
\author[1]{William J. McCausland}
\affil[1]{D\'epartement de sciences \'economiques and CIREQ\\Universit\'e de Montr\'eal}

\date{\today}

\begin{document}

\maketitle

\begin{abstract}
\noindent
We introduce a new stochastic duration model for transaction times in asset markets.
We argue that widely accepted rules for aggregating seemingly related trades mislead inference pertaining to durations between unrelated trades: while any two trades executed in the same second are probably related, it is extremely unlikely that {\em all} such pairs of trades are, in a typical sample.
By placing uncertainty about which trades are related within our model, we improve inference for the distribution of durations between unrelated trades, especially near zero.
We introduce a normalized conditional distribution for durations between unrelated trades that is both flexible and amenable to shrinkage towards an exponential distribution, which we argue is an appropriate first-order model.
Thanks to highly efficient draws of state variables, numerical efficiency of posterior simulation is much higher than in previous studies.
In an empirical application, we find that the conditional hazard function for durations between unrelated trades varies much less than what most studies find.
We claim that this is because we avoid statistical artifacts that arise from deterministic trade-aggregation rules and unsuitable parametric distributions.

\vspace*{0.65\baselineskip}
\end{abstract}
\clearpage

\section{Introduction}\label{s:intro}
Duration models for financial transactions describe the irregular timing of trades, or other events such as price changes.
They are useful because trading intensity is one measure of market liquidity.
And unlike count models, duration models use all the information in trading times.
They also shed light on market microstructure phenomena.

Modelers have to confront the fact that some trades are related to others and that related trades are nearly simultaneous.
We will argue that widely accepted rules for aggregating seemingly related trades into clusters, together with unsuitable parametric duration distributions, mislead inference in important ways.
The most commonly used rule treats two trades as related if they are executed, to within recording precision, simultaneously.
In this paper, as in most studies, transaction times are truncated to the second, and in this context we will call it the {\em same-second} aggregation rule.\footnote{Many datasets, especially more recent ones, feature millisecond or finer recording precision. The issues raised in this paper apply widely, but since the unfortunate consequences of deterministic aggregation rules are qualitatively different when recording precision is finer, we leave this case for future research.}
The rule is usually correct, case by case, but in a reasonably sized sample from a liquid market, it is likely that many pairs of unrelated trades occur within the same clock second by happenstance.

Our paper makes three main contributions.
First, we put into our model the uncertainty about whether each duration is a {\em cluster} duration---one between related trades---or a {\em regular} duration---one between unrelated trades.
Not surprisingly, the posterior probability that a duration recorded as $0s$ (zero seconds) is due to coincidence, and therefore regular, is always low.
But it is never zero, and the posterior mean of the number of these coincidences is much larger than its posterior standard deviation.
We find that the conditional hazard function for regular durations varies smoothly near zero, in contrast to the abrupt changes found in most studies, which we argue are artifacts arising from classifying {\em all} durations recorded as $0s$ as cluster durations.

Our second contribution is a new flexible model for regular durations.
We identify some undesirable features of existing parametric conditional duration densities and propose a new family of conditional distributions that is flexible, but also amenable to shrinkage towards the exponential distribution, which we will argue is a theoretically appealing first-order model.

Our third contribution is computational, and promotes highly numerically efficient posterior simulation.
We use the HESSIAN method, introduced by \cite{McCa12}, to draw, in a single Gibbs block, the full sequence of state variables describing trade intensity, together with various parameters.
To date, the HESSIAN method has only been applied to non-Gaussian state-space models with parametric distributions for observed variables and homogenous state transitions.
Here we use it in a model with a flexible distribution for observables and heterogenous state transitions.
The numerical efficiency we achieve is in fact considerably higher than that achieved using auxiliary mixture model methods, which rely on special features of parametric distributions.

Two other features of our model are appealing and uncommon, but not original.
First, we estimate the regular diurnal (time of day) pattern of trading intensity jointly with other features of the model.\footnote{
\cite{VereRodrEspa02} and \cite{BrowVann13} jointly estimated the diurnal patterns and the parameters for ACD models using respectively a semi-parametric approach and MCMC methods within a Bayesian framework.
}
Second, the latent state process is an irregularly sampled Ornstein-Uhlenbeck (OU) process, rather than a homogenous autoregressive process; autocorrelations depend on the elapsed time, rather than the number of intervening trades, between two durations.

We now illustrate in detail the problem with treating all durations recorded as $0s$ as cluster durations, a practice that goes back to the seminal paper of \cite{EnglRuss98}.
Table \ref{tab:descriptive-statistics} shows descriptive statistics for two duration series, cleaned as described in Section \ref{sec:data}.
Recorded durations are differences between two trading times, both truncated to the second.
The last six columns give the percentage of durations recorded as $0s$, $1s$, $2s$, $3s$, $4s$, and $5s$.
The values from $1s$ to $5s$ vary smoothly, but far more durations are recorded as $0s$ than are recorded as $1s$.

This zero inflation clearly needs to be addressed.
It arises because of nearly simultaneous related trades.
These often occur when a market order is matched against, and filled with, several limit orders on the opposite side of the market.
It may also occur if many traders submit limit orders to be executed at a round price, as suggested by \cite{VereRodrEspa02}, or if important news synchronizes a flurry of trading, or if traders use algorithmic trading strategies that can be triggered by another trade.
We do not try to distinguish different causes of related trades.

In fact, the amount of zero inflation is greater than it might appear: the duration between two trades will be recorded as $0s$ if the second trade occurs during the {\em remainder} of the same clock second as the first, but as $1s$ if it occurs at {\em any} time during the next clock second.
Extrapolating percentages from $1s$ to $5s$ back to $0s$ and dividing by two gives us a rough idea of the percentage of all durations that are regular and recorded as $0s$: about 5\% for the RY series and 7\% for the POT series.
Of course, these represent a much larger percentage of regular durations, around 14\% and 22\%, respectively.
While it is likely that a large majority of durations recorded as $0s$ are cluster durations, it would be very surprising if all were.

The same-second aggregation rule amounts to removing all durations recorded as $0s$ from the sample.
The result is a truncated sample of regular durations.
The fact that the truncation is at both the mode and lower bound of the distribution is particularly unfortunate.
Trading intensity is understated, and the understatement varies with trading intensity: when it is high, there are more short regular durations and more spurious aggregation of unrelated trades.

\begin{figure}[t]
    \centering
    \begin{subfigure}{0.475\textwidth}
    	\centering
    	\includegraphics[width=\textwidth]
    	{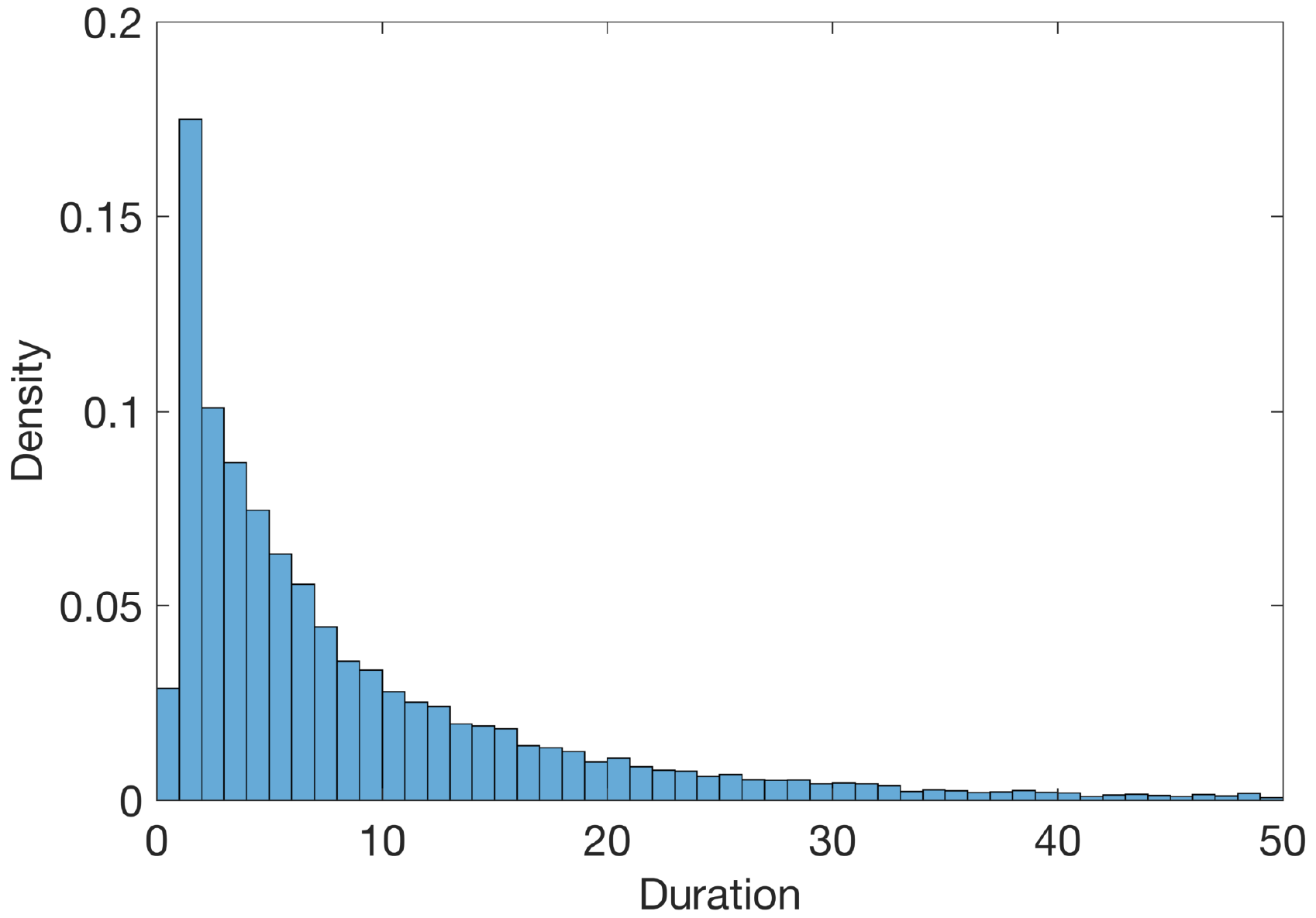}
    \end{subfigure}
    \quad
    \begin{subfigure}{0.475\textwidth}  
    	\centering 
    	\includegraphics[width=\textwidth]
    	{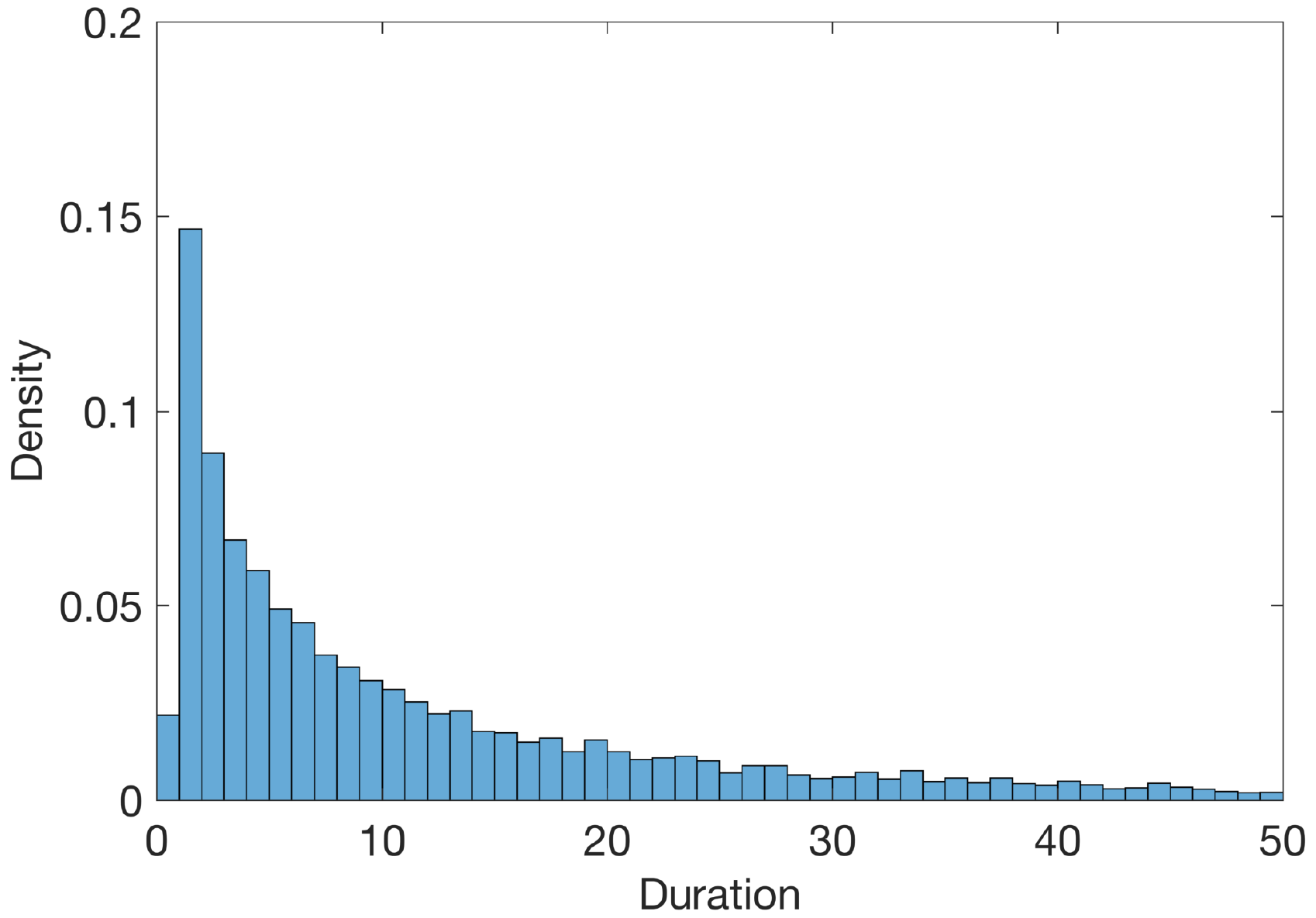}
    \end{subfigure}
	\caption
	{\small
	Histograms of regular durations, as classified by the GW aggregation rule, from $0s$ to $50s$, for the RY series on the left and the POT series on the right. Bins are aligned with clock seconds.
	}
	\label{fig:aggregation}
\end{figure}

Another aggregation rule, proposed by \cite{GramWell02}, aggregates any sequence of transactions within the same clock second where prices are non-decreasing or non-increasing.
We will call this the {\em GW} aggregation rule.
Figure~\ref{fig:aggregation} shows histograms of durations that are classified as regular by this rule.
Not enough of the durations recorded as $0s$ are classified as regular for compatibility with a smooth density near zero.
This is not surprising, as many pairs of unrelated trades will feature price changes of the same sign by coincidence.

While the spurious aggregation of unrelated trades is our main concern with these rules, we also question whether even error-free aggregation would be desirable.
If a market order matches nine limit orders, five times as many traders are getting their orders filled than if it matches a single limit order.
Treating these cases as equivalent may understate liquidity, as perceived by traders.

Our model is an example of what \cite{Engl02} calls a {\em multiplicative error} model, where the scale of the conditional duration distribution depends on the history of the process, as well as observables such as time of day, but the shape does not.\footnote{
For recent surveys on the analysis of high-frequency financial durations using multiplicative error models see \cite{Pacu08}, \cite{Haut12} and \cite{BhogVari18}.
}
We will call the conditional distribution of duration divided by scale the {\em normalized} conditional distribution; typically, it has unit mean.
Two basic models are the data-driven Autoregressive Conditional Duration model (ACD), introduced by \cite{EnglRuss98}, where the scale depends deterministically on past durations; and the parameter-driven Stochastic Conditional Duration model (SCD) introduced by \cite{BauwVere04}, where it is a latent stochastic process.
\cite{BauwGiot00} propose a logarithmic version of the ACD model, avoiding parameter restrictions; SCD models usually feature a similar logarithmic specification.
\cite{BauwVere04} provide empirical evidence favouring the SCD model over a similar ACD model in the four data sets they analyzed.

The most commonly used conditional distributions in SCD and ACD models are the exponential, and two generalizations: the gamma and the Weibull \citep{EnglRuss98, BauwGiot00, BauwVere04, FengJianSong04, StriForbMart06, MenKolkWirj15}.
\cite{EnglRuss98} and others conclude that the exponential is too inflexible and favour the Weibull or gamma.
However, these conclusions favouring the Weibull or gamma come largely from studies in which durations recorded as $0s$ are removed from the sample, and the density and hazard functions of these distributions feature extreme behaviour near zero.
Except for the knife-edge special case where they reduce to an exponential, which has a constant  hazard function, the hazard function  is either zero at a duration of zero and increasing; or infinite at a duration of zero and decreasing.\footnote{
The generalized gamma and the Burr distribution generalize the gamma and Weibull distribution.
They were proposed as conditional distribution for ACD models by \cite{Lund99} and \cite{GramMaue00}.
These distributions, unlike the gamma and the Weibull, allow for non-monotonic hazard functions, but they retain the property that their hazard function is bounded away from zero and infinity only for very special cases.}

We argue that extreme variation of the hazard function of regular durations near zero is implausible.
In queueing theory, a simple model for arrival times (of, say, customers at an ATM) is the Poisson process.
It is reasonable when there are a large number of potential customers, acting independently and homogenously in time, and the probability of any {\em given} customer arriving in a given time interval is much smaller than the probability of {\em some} customer arriving in that interval.
In a Poisson process, durations between arrivals are exponentially distributed.
The constant hazard function of the exponential makes it {\em memoryless}: the probability of an arrival in the next minute does not depend on how long you have been waiting.

Of course, trading intensity in financial markets varies over time.
But after conditioning on relevant predictors and latent states measuring trading intensity, we would expect the distribution of regular durations to be not far from an exponential---its hazard rate a function of this conditioning information---due to the large number of unrelated potential traders and the fact that a regular duration is, by definition, the time interval between unrelated trades.

The hazard functions of mixtures of exponentials are bounded away from zero and infinity,\footnote{ \cite{DeLuGall04} used a mixture of two exponentials in ACD models and found that this specification provides a better fit than a Weibull distribution.
\cite{DeLuGall09} again use a mixture of two exponentials but allow mixture weights to depend on observable market activity.
}
but they are decreasing---see \cite{BarlMarsPros63}---which is a restrictive feature.
Moreover, not all decreasing hazard functions can be easily captured by mixtures of exponentials.
In simulations we do not report, we find that adding mixture components after the second yields little: only components with the largest and smallest hazards are important, in the sense that the posterior distributions of the weights of other components are highly concentrated near zero.\footnote{ Other mixture distributions have been proposed for the normalized conditional density.
\cite{WirjKolkMen13} did a Bayesian analysis of the SCD model, with leverage, using three types of two-component mixtures: two exponentials, two Weibulls and two gammas.
}

The rest of the paper is organized as follows.
We describe the transaction data we analyze in Section~\ref{sec:data}, our SCD model in Section~\ref{sec:model} and our methods for posterior simulation in Section~\ref{sec:estimation}.
In Section~\ref{sec:results}, we conduct an artificial data experiment to test for the correctness of our posterior simulator and then illustrate our methods in an application featuring two equities traded on the Toronto Stock Exchange.
We conclude in Section~\ref{sec:conclusion}.

\section{Data}\label{sec:data}
Our data comes from the Tick Data historical database of securities listed on major stock exchanges.
We use transaction data for two equities traded on the Toronto Stock Exchange (TSX): the Royal Bank of Canada (RY), and the Potash Corporation of Saskatchewan (POT).
For each equity, we observe transactions over five consecutive trading days, from March 17 to March 21, 2014.

The trading hours of the TSX are 9:30 am to 4:00 pm, Monday to Friday.
We ignore the {\em pre-open} trading session and the {\em extended} trading session after closing, and only use data from the {\em continuous} trading sessions.
These data include the time of each transaction (the time stamp), the price, the volume (in number of shares), the type of trading session (pre-open, continuous, or extended), and various indicator variables specifying, among other things, whether the transaction was delayed or subsequently corrected.
Trading times are truncated to the second; in other data sets, transaction times are recorded to higher levels of precision.

\begin{table}[t]
\begin{center}
\begin{threeparttable}
    \caption
    {\small
    Descriptive statistics of the cleaned sample from March 17 to 21, 2014.
    }
    \label{tab:descriptive-statistics}
    \begin{tabular}{l c c c c c c c c c c c c }
    	\toprule
         & Trades & Mean & Std. & Max & C.V. &
        $0s\,\%$ & $1s\,\%$ & $2s\,\%$ & $3s\,\%$ & $4s\,\%$ & $5s\,\%$ \\
        \midrule
        \input{descriptive-statistics.tex}
        \bottomrule
    \end{tabular}
    {\footnotesize
    The first column gives the number of transactions.
    The next four columns give the mean, standard deviation, maximum value and coefficient of variation of durations.
    The last six columns give the percentage of trade durations recorded as $0s$, $1s$, $2s$, $3s$, $4s$, and $5s$.
    }
\end{threeparttable}
\end{center}
\end{table}

\cite{BrowGall06} argue that careful data cleaning is an important part of high-frequency duration analysis.
We follow standard practice to remove observations that are obviously erroneous or result from atypical market conditions, such as at opening and closing.
We delete entries identified by the exchange as incorrect, corrected, delayed or canceled.
Then we remove transactions with aberrant prices using the method suggested by \cite{BrowGall06}.
Descriptive statistics of the cleaned data are reported in Table~\ref{tab:descriptive-statistics}.
For each duration series, we report the number of observations, followed by the sample mean, standard deviation, maximum, and coefficient of variation.
An exponential has a coefficient of variation of 1, so the empirical distributions are considerably overdispersed relative to the exponential.
The second part of the table reports the percentage of durations recorded as $0s$, $1s$, $2s$, $3s$, $4s$, and $5s$.

\begin{figure}[t]
    \centering
    \begin{subfigure}{0.475\textwidth}
    	\centering
    	\includegraphics[width=\textwidth]
    	{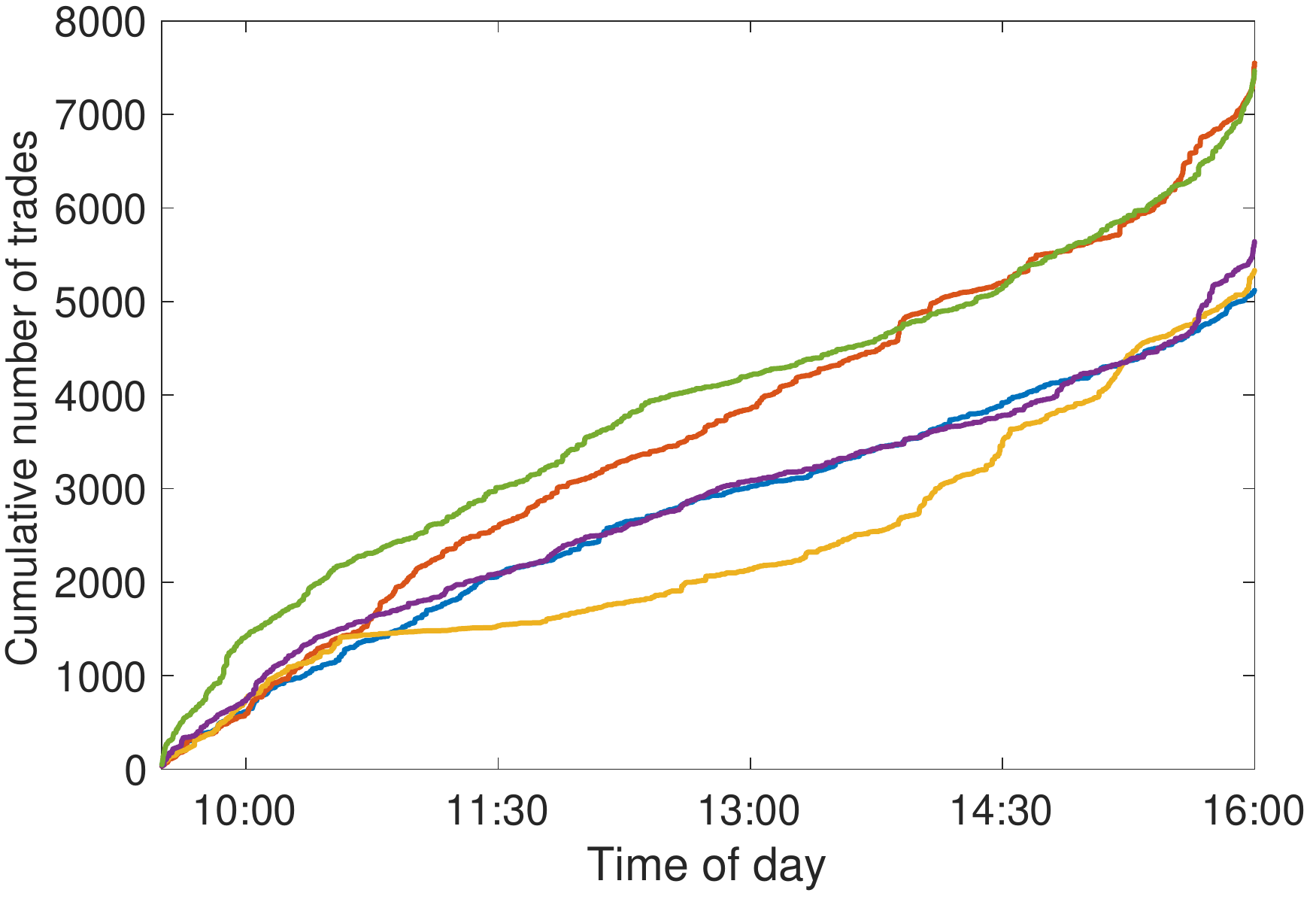}
    \end{subfigure}
    \quad
    \begin{subfigure}{0.475\textwidth}  
    	\centering 
    	\includegraphics[width=\textwidth]
    	{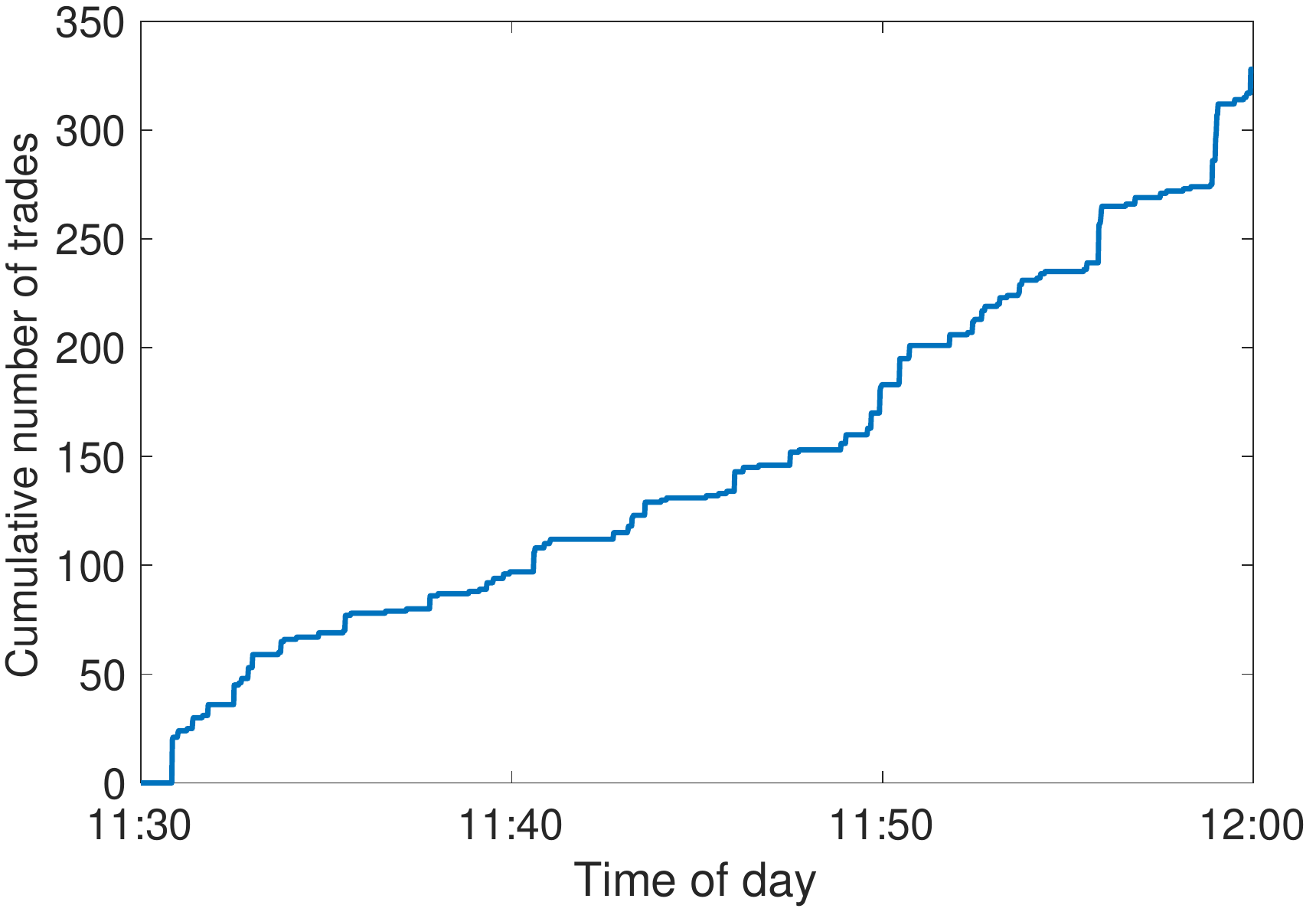}
    \end{subfigure}
	\caption
	{\small
	Cumulative numbers of trades for the POT series; on the left, for each day from March 17 to 21, 2014; on the right, for the interval between 11:30am and noon on March 17, 2014.
	}
	\label{fig:cumulative}
\end{figure}

The left panel of Figure~\ref{fig:cumulative} shows the cumulative number of trades of POT during each of five consecutive days of trading, from March 17 to 21, 2014.
The right panel zooms in on a typical half-hour period to give finer detail; it shows the cumulative number of trades on March 17, 2014 between 11:30am and noon.
The figure illustrate many widely known stylized facts about durations.
Trading intensity varies over time.
Part of the variation is predictable given time of day: there are more trades near the beginning and end of the trading session than in the middle.
The remaining, stochastic, part is highly persistent.
Trades often arrive in clusters, within a very short interval and without a marked difference in trading intensity before and after the cluster.

\section{A Stochastic Conditional Duration Model}\label{sec:model}
Our model builds on the multiplicative-error stochastic volatility model of \cite{BauwVere04}.
It differs in three important ways.
First, it has a second state variable, one which indicates which durations are cluster durations and which are regular.
Since this replaces the usual practice of aggregating trades into clusters before analysis, our model is intended for unaggregated data.
For the purpose of comparison, we also define a special case of our model where cluster durations have probability zero, for samples in which durations have been aggregated into clusters.
We will call the special case the {\em regular-duration} model and the general model the {\em all-duration} model.
Placing the classification of trades within the model makes it possible to infer that the number of regular durations that happen to be recorded as $0s$, while small, is not zero.
We will see that this overcomes the problems arising from the spurious aggregation of unrelated trades.
Second, we use a flexible distribution for the scale-normalized duration distribution.
Our approach makes it easy to choose a prior distribution for this distribution that shrinks towards an exponential distribution, which we have argued is a sensible first-order conditional model for regular durations.
Third, we jointly estimate the diurnal (time-of-day) pattern along with everything else, which gives a better accounting of the uncertainty about how trading intensity evolves.
We describe a family of prior distributions to complete the model.

As with most models in the literature, time is continuous, which may not be suitable for data where trading intensity is high and transaction times are truncated to the second.
In Section~\ref{sec:estimation-adjustment} below, we describe a discrete-time version of the model that is more suitable for such data.

\subsection{The Data Generating Process}\label{sec:dgp}
We observe transaction times over $D$ trading days in the interval $[\topen, \tclose]$, where
$\topen$ and $\tclose$ are the opening and closing times.
All times of day are measured in seconds after midnight.
For each day $d=1,\dots,D$, denote the sequence of transaction times by $t_{d,0},t_{d,1},\dots,t_{d,n_d}$ and construct the durations $\ydi \equiv \tdi - \tdim$, $i=1,\dots,n_d$.

The conditional distribution of each $\ydi$ depends on the values $\sdi \in \{0,1\}$ and $\xdi \in \mathbb{R}$, at time $\tdim$, of two latent states.
The state $\sdi$ is a mixture component indicator: $\sdi = 0$ indicates that $\ydi$ is a cluster duration; $\sdi = 1$, a regular duration.
The state $\xdi$ gives the trading intensity at $\tdim$, but only if the duration is regular.
The conditional density of $\ydi$, given $\sdi$ and $\xdi$, is
\begin{equation}\label{eq:mix-density}
	p(\ydi \vbar \sdi, \xdi) = 
	\begin{cases}
	p_0(\ydi)				& \quad \sdi = 0, \\
	p_1(\ydi \vbar \xdi)	& \quad \sdi = 1,
	\end{cases}
\end{equation}
where $p_0$ and $p_1$ are density functions on $[0, \infty)$.

At each day $d$, the indicator process $\sdi$ is first-order Markov and stationary, with 
\begin{equation}\label{eq:markov-prob}
	\Pr[\sdip = l \vbar \sdi = k] = \xi_{kl} 
	\quad \text{and} \quad 
    \Pr[\sdf = k] = \xi_{k}, 	
\end{equation}
where $\xi_0 \equiv (1-\xi_{11}) / (2-\xi_{00}-\xi_{11})$ and $\xi_1 \equiv (1-\xi_{00}) / (2-\xi_{00}-\xi_{11})$ by stationarity.
The trading intensity process $\state_d(t)$, $t \in [\topen,\tclose]$, is the sum of a deterministic function $\mean(t)$ describing a diurnal pattern---equation \eqref{eq:diurnal-function} below---and a zero-mean OU process.
Sampling the $\state_d(t)$ process at all transaction times gives the values $\xdi \equiv \state_d(\tdim)$; then the discrete time process $\xdi$ is first order autoregressive, but not homogenous due to irregularly spaced trading times:
\begin{align}\label{eq:dynamic-OU}
\begin{split}
 	\xdip \vbar \xdi,\tdi,\tdim
 	&\sim \N
 	\left(
 	\mean(\tdi) + e^{-\phi \ydi}(\xdi - \mean(\tdim)),\sigma^2 (1-e^{-2\phi \ydi})
 	\right),\\
 	\xdf \vbar \tdf
 	&\sim \mathcal N
 	\left(
 	\mean(\tdf), \sigma^2
 	\right), 	
\end{split}
\end{align}
where $\phi \geq 0$ is the mean reversion parameter, and $\sigma$ is the marginal standard deviation parameter of the OU process.

To model the diurnal pattern, we follow \cite{EileMarx96}, \cite{LangBrez04} and \cite{BrowVann13} and specify the deterministic process $\mean(t)$ as a cubic B-spline function, a piecewise polynomial defined on a set of $M$ equally spaced knots $\topen = \kappa_1 < \cdots < \kappa_M = \tclose$.
Knots $\kappa_1$ and $\kappa_M$ have multiplicity 4 and the rest have multiplicity 1.
This gives an expansion that is the following linear combination of $L=M+2$ B-spline basis functions:
\begin{equation}\label{eq:diurnal-function}
	\mean(t) = \sum_{l=1}^{L} \delta_l B_{l}(t),	
\end{equation}
where $B_l(\cdot)$ denotes the $l$-th B-spline basis function, a local cubic polynomial, and $\delta_l$ its coefficient.
The basis functions depend on the location and multiplicity of the knots.
Setting the multiplicity of the knots $\kappa_1$ and $\kappa_M$ to 4 makes $\mean(\topen) = \delta_1$ and $\mean(\tclose) = \delta_L$.
This makes $\delta_1$ and $\delta_L$ easier to interpret.
See \cite{Boor_1978} or \cite{Dierckx_1993} for more on B-splines.

We now specify the component densities $p_0(\cdot)$ (cluster) and $p_1(\cdot)$ (regular).
We specify $p_0(\cdot)$ as the following mixture of two exponentials, with hazard parameters $\lambda_1$ and $\lambda_2$:
\begin{equation}\label{eq:cluster}
	p_0(\ydi) = 
		\pi \lambda_1 e^{-\lambda_1 \ydi} + (1-\pi) \lambda_2 e^{-\lambda_2 \ydi}.
\end{equation}
Given the nature of cluster durations, $p_0(\cdot)$ should concentrate most of its probability near zero; the prior distribution over $\lambda_1$, $\lambda_2$ and $\pi$ should be chosen so that this is true with high prior probability.

Unlike cluster durations, regular durations depend on market conditions, as captured by the intensity state $\xdi$.
Our specification of $p_1(\cdot)$ conforms to the multiplicative error structure proposed by \cite{BauwVere04}, which gives the regular duration as $\ydi = e^{\xdi} \epsilon_{d,i}$,
where $\epsilon_{d,i}$ is the contemporaneous value of a non-negative iid process with $E[\epsilon_{d,i}] = 1$.
The $\epsilon_{d,i}$ and $\xdi$ processes are independent.
We denote the density of $\epsilon_{d,i}$ by $p_\epsilon(\cdot)$ and call it the normalized duration density.
Thus, the conditional density of regular duration $\ydi$, given $\xdi$, is
\begin{equation}\label{eq:regular}
	p_1(\ydi \vbar \xdi) = e^{-\xdi} p_\epsilon \left(\ydi e^{-\xdi}\right).
\end{equation} 
Given $\sdi=1$, the quantity $e^{\xdi}$ is the conditional mean of $\ydi$.
It gives only the scale of the distribution, the shape being determined by $p_\epsilon(\cdot)$.
We argued that some commonly used duration distributions are unsuitable because of their restrictive or implausible hazard functions.
To complete the model specification, we will now propose a new flexible family of normalized densities.

\subsection{A Normalized Density for Durations}\label{sec:density}
We adopt an approach similar to one described in \cite{FerrStee06}, where the cdf $P_\epsilon(\cdot)$ of a flexible density $p_\epsilon(\cdot)$ is constructed as
$
    P_\epsilon(\epsilon) \equiv G\big(F(\epsilon)\big),
$
where $F(\cdot)$ is a parametric continuous cdf, with density $f(\cdot)$, having the same support as $P_\epsilon(\cdot)$; and $G(\cdot)$ is a flexible continuous cdf on $[0,1]$, with density $g(\cdot)$.
The cdf $P_\epsilon(\cdot)$ can be viewed as a perturbed version, depending on $G(\cdot)$, of the parametric cdf $F(\cdot)$.
When $G(\cdot)$ is uniform on $[0,1]$, there is no perturbation and $P_\epsilon(\cdot) = F(\cdot)$.
As noted by \cite{FerrStee06}, distributions defined in this way cover the entire class of continuous distributions, since any such $P_\epsilon(\cdot)$ can be constructed for a suitable choice of $G(\cdot)$.
The construction implies the normalized density
\begin{equation}\label{eq:transform}
    p_\epsilon(\epsilon) = f(\epsilon)g\big(F(\epsilon)\big).
\end{equation}

We argued that the exponential distribution was a theoretically promising first order approximation of a conditional duration distribution, and so we specify $F(\cdot)$ as an exponential, in the hope of capturing realistic hazard functions using a distortion $G(\cdot)$ with a small number of terms.
The hazard rate of $F(\cdot)$, which we denote $\tilde \lambda$, will be substituted out using the scale normalization condition $E[\epsilon] = 1$.
At the same time, we want $P_\epsilon(\cdot)$ to be flexible, allowing for large departures from the exponential if the data warrant it.
For this reason, we choose a flexible functional form for $g(\cdot)$, a $J$'th order Bernstein polynomial with positive coefficients adding to one.
This is a $J$-component mixture of beta densities, each with integer-valued shape parameters adding to $J+1$; coefficients of the first and last terms determine $g(0)$ and $g(1)$, respectively.
Bernstein polynomials can approximate any continuous density on $[0,1]$ arbitrarily closely (in $\sup$ norm) for $J$ sufficiently large.\footnote{
	Let $F(\cdot)$ be a bounded continuous function on $[0,1]$.
	The Bernstein polynomial of order $n$ of the function $F(\cdot)$ is defined as
	\[
		B_n^F(x) = \sum_{k=0}^n F\left(\frac{k}{n}\right)\binom{n}{k} x^k (1-x)^{n-k}.
	\]	
	\cite{Lore53} shows that $\lim_{n\rightarrow\infty} \sup_{x \in [0,1]} |B_n^F(x) - F(x)| = 0$.
}
The component probabilities are the coefficients of the Bernstein basis polynomials.
Specifically,
$
    g(z) =  \sum_{j=1}^J \beta_j \Be(z \vbar j,J-j+1),
$
where $\sum_{j=1}^J \beta_j = 1$, $\beta \equiv (\beta_1,\ldots,\beta_J) \geq 0$, and $\Be(z \vbar a, b)$ denotes the beta density with shape parameters $a$ and $b$, for $a,b > 0$.
The $J$ Bernstein basis polynomials of order $J$ form a partition of unity, so that if $\beta = (1/J,\ldots,1/J)$, then $g$ is uniform and we get back the original exponential density, $p_e(\cdot) = f(\cdot)$.
This makes it easy to choose a prior distribution for $\beta$ so as to centre the induced prior for $p_\epsilon(\cdot)$ around the exponential distribution and to control the amount of shrinkage towards it.
We treat the order $J$ of the Bernstein polynomial as fixed, not as a parameter to estimate, due to the difficulty of computing marginal likelihoods for models with two different state variables.
Instead of estimating $J$, we will compare results over different values of $J$.
For a discussion of Bayesian nonparametric density estimation using Bernstein polynomials with unknown $J$, see \cite{Petr99a,Petr99b} and \cite{PetrWass02}.

Substituting the exponential distribution function $F(\epsilon) = 1-e^{-\tilde \lambda \epsilon}$ and the above expression for $g(z)$ into equation~\eqref{eq:transform} gives $p_\epsilon(\cdot)$ as a polynomial in $e^{-\tilde \lambda \epsilon}$, which we write explicitly as the following linear combination of exponential densities:
\begin{equation}\label{eq:linear-exp}
  p_\epsilon(\epsilon) = \sum_{j=1}^J \alpha_j j\tilde \lambda e^{-j\tilde \lambda \epsilon},
\end{equation}
where $\alpha \equiv (\alpha_1,\dots,\alpha_J) = T\beta$ and $T$ is a $J\times J$ matrix with $T_{j,J-i+1} = \binom{J}{j}\binom{j}{i}(-1)^{j-i} \, (i/j)$ if $j\leq i$ and $0$ otherwise.
Non-negativity of $\beta$ ensures the non-negativity of $p_\epsilon(\cdot)$, but as the $\alpha_j$ are not necessarily all non-negative, $p_\epsilon(\cdot)$ is not necessarily a mixture of exponentials.
It is easy to check that the hazard function for $p_\epsilon(\cdot)$ is bounded away from zero and infinity, and its limiting value as $\epsilon$ goes to infinity is the hazard parameter $\tilde \lambda$ of the exponential distribution with cdf $F(\cdot)$.
The scale normalization condition $\E[\epsilon]=1$ gives
\(
    \tilde \lambda = \sum_{j=1}^J \frac{\alpha_j}{j},
\)
which we use to substitute out $\tilde \lambda$ from the expression for $p_\epsilon(\cdot)$, freeing us from having to impose restrictions on $\alpha$ or $\beta$.

\begin{figure}[t]
	\centering
    \begin{subfigure}{0.475\textwidth}
    	\centering
    	\includegraphics[width=\textwidth]
    	{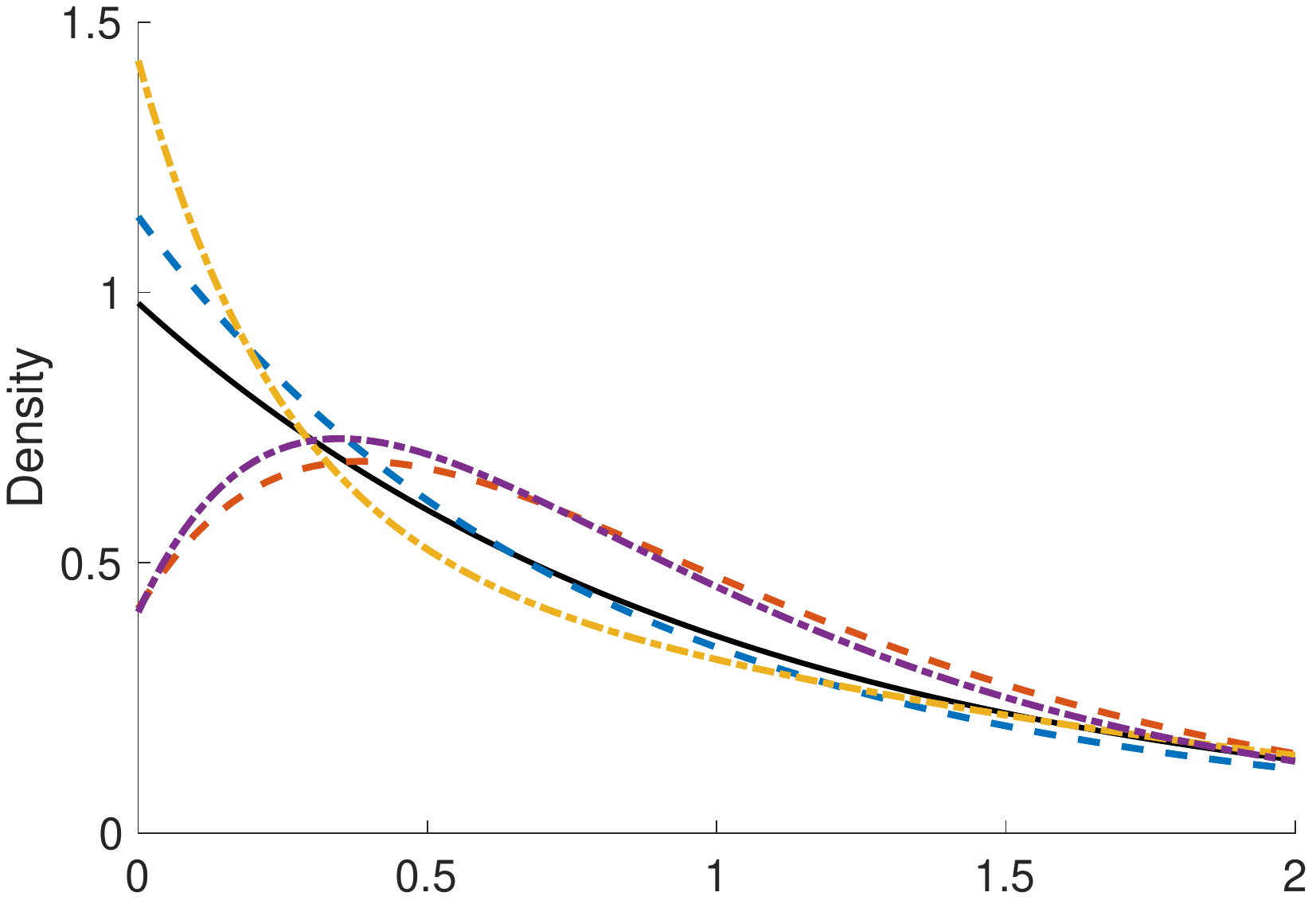}
    \end{subfigure}
    \quad
    \begin{subfigure}{0.475\textwidth}  
    	\centering 
    	\includegraphics[width=\textwidth]
    	{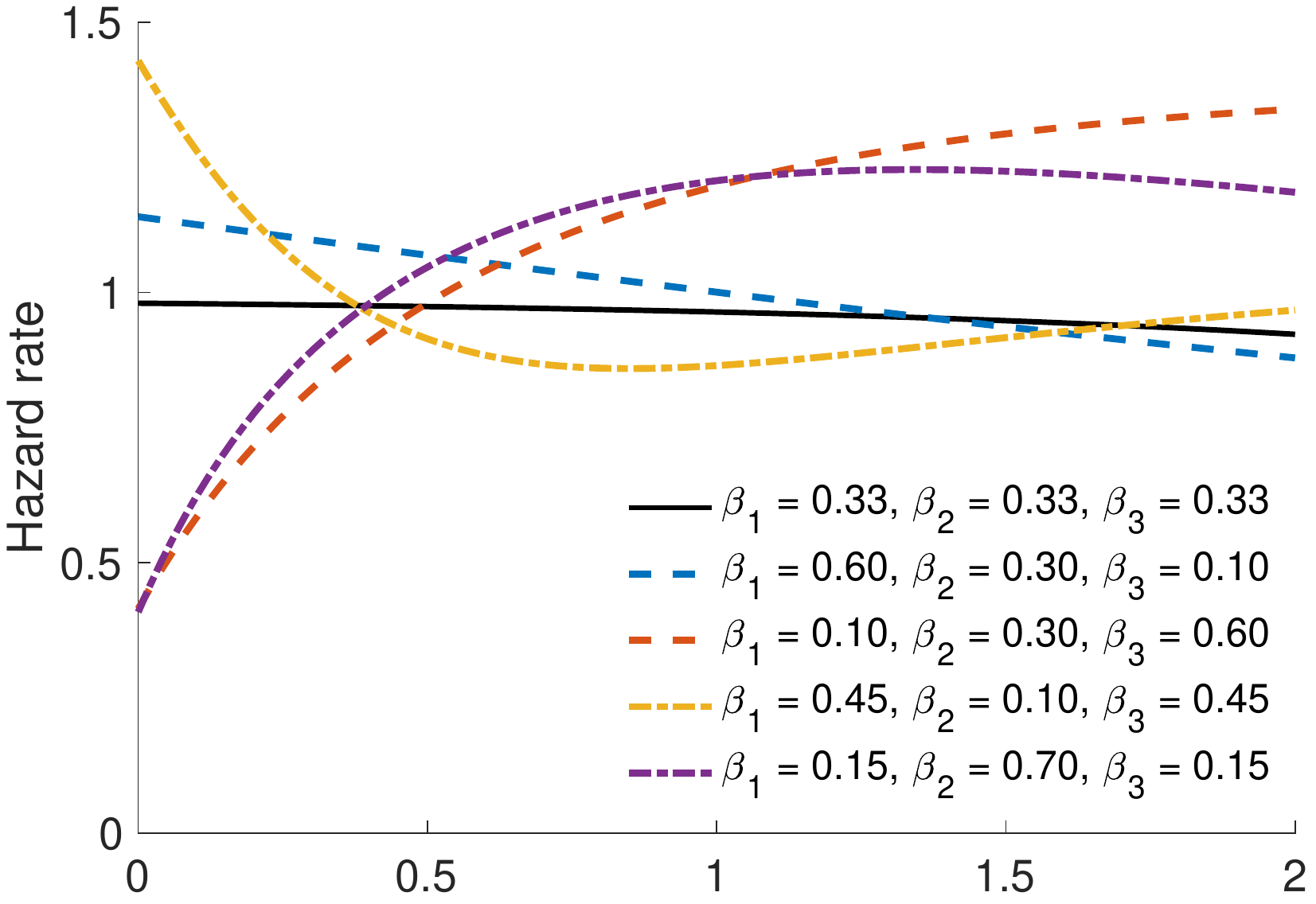}
    \end{subfigure}
	\caption
	{\small
	Density and hazard functions for a unit-mean exponential and five distributions with $J=3$ terms.
 	}
	\label{fig:examples-density-harzard}
\end{figure}

We now illustrate the flexibility we achieve with just a few terms.
Figure~\ref{fig:examples-density-harzard} shows examples of decreasing, increasing and non-monotonic hazard functions that can be captured with $J=3$ components.
The solid lines are the density and (constant) hazard functions of an exponential with mean equal to one.
The other density and corresponding hazard functions are for various values of $(\beta_1,\beta_2,\beta_3)$.
Taking into account the adding-up constraint, there are two degrees of freedom, the same as a  unit-mean mixture of two exponentials.
The dashed lines show pairs of density and hazard functions where the hazard is monotonic; dash-dotted lines, pairs where it is not monotonic.
Recall that all mixtures of exponentials have a decreasing hazard function.

\subsection{Prior Distributions}
To complete the model, we describe a prior distribution for the parameter vectors $(\phi,\sigma)$, $\delta$, $\beta$, $\xi$, $\pi$ and $\lambda$.
These vectors are {\em a priori} independent.

We specify a multivariate log-Normal prior distribution for the vector $(\phi,\sigma)$ of OU parameters:
$
	\theta \equiv (\log \phi,\log \sigma) \sim \mathcal N(\bar \theta, \bar \Sigma).
$
The transformation eliminates the need for parameter restrictions.

We induce a Gaussian prior for the vector $\delta$ of coefficients defining the diurnal pattern by specifying the following Gaussian prior on an invertible linear function of $\delta$:
\begin{equation}\label{eq:deltainduce}
  \begin{bmatrix} v' \\ \Delta \end{bmatrix} \delta
  \sim
  \N\left(
    \begin{bmatrix} \bar{\delta} \\ 0_{L-1} \end{bmatrix},
    \begin{bmatrix} \bar{h} & 0_{L-1}' \\ 0_{L-1} & \tau I_{L-1} \end{bmatrix}^{-1}
  \right),
\end{equation}
where $v$ and $\Delta$ are the mean and first difference operators defined by
\[
  v \equiv \frac{1}{L} \begin{bmatrix} 1 \\ \vdots \\ 1 \end{bmatrix},
  \quad
  \Delta \equiv \begin{bmatrix} -1 & 1 & 0 & \hdots & 0 \\ 0 & -1 & 1 & \ddots & \vdots \\ \vdots & \ddots & \ddots & \ddots & 0 \\ 0 & \hdots & 0 & -1 & 1 \end{bmatrix},
\]
$0_{L-1}$ is an $(L-1) \times 1$ vector of zeros, $I_{L-1}$ is the $(L-1) \times (L-1)$ identity matrix, $\bar{\delta}$ is a scalar location parameter and $\tau$ and $\bar{h}$ are scalar precision parameters.
Thus, first differences $\delta_l - \delta_{l-1}$ are iid $\N(0,\tau^{-1})$ and the arithmetic mean $v\delta$, is independent of the first differences, with $v\delta \sim \N(\bar{\delta}, \bar{h}^{-1})$.
This is the same as the random walk prior proposed by \cite{LangBrez04}, except that they use $v = (1, 0, \ldots, 0)$.
Both priors favour smoothness and are agnostic with respect to the signs of derivatives; since the derivatives of B-splines are linear combinations of the first differences $\delta_l - \delta_{l-1}$, we can interpret $\tau$ as a smoothing parameter for the diurnal pattern.
Higher values of $\tau$ imply more shrinkage towards a flat diurnal pattern.
Unlike the prior proposed by \cite{LangBrez04}, where the prior variances of the $\delta_l$ increase with $l$, our prior has the symmetry property that $(\delta_1,\ldots,\delta_L)$ and $(\delta_L,\ldots,\delta_1)$ have the same distribution, and prior uncertainty about trading intensity is the same at the opening and at the closing of the market.

Equation \eqref{eq:deltainduce} induces the prior $\delta \sim N(\bar{\mu}, \bar{H})$, where $\bar{\mu} = \bar{\delta} L v$ and $\bar{H} = \bar{h} vv' + \tau \Delta' \Delta$.
The prior covector, which we use in Section \ref{sec:estimation} below, is $\bar{c} \equiv \bar{H}\bar{\mu} = \bar{h} v$.
We will estimate $\tau$, and specify its prior as the following scaled chi-square: $\bar s \tau \sim \chi^2(\bar \nu)$.

For the beta mixture weights indexing the normalized duration density, we specify $\beta \sim \Dir(\bar M \bar \beta)$,
  where $\bar \beta=(\bar \beta_1,\dots,\bar \beta_J) > 0$, $\sum \bar \beta_j=1$, and $\bar M>0$.
The prior mean of $\beta$ is $\bar \beta$; $\bar M$ is a concentration parameter.
When $\bar \beta = (1/J,\dots,1/J)$, the prior mean of $\beta$ corresponds to $g(\cdot)$ being uniform on $[0,1]$ and therefore $p_\epsilon(\cdot)$ being an exponential density.
In this sense, $\bar \beta = (1/J,\dots,1/J)$ centres the prior distribution for $p_\epsilon(\cdot)$ around the exponential density, with $\bar M$ controlling the amount of shrinkage towards it; higher values of $\bar M$ imply more shrinkage.

The diagonal elements of the Markov transition matrix $\xi$ of the latent indicator $\sdi$ are independent and beta distributed: $ \xi_{kk} \sim \Be(\bar a_k,\bar b_k)$, $k=0,1$.
The hazard parameters of the cluster duration density are independent gammas: $\lambda_k \sim \Gam(\bar a_{\lambda_k}, \bar b_{\lambda_k})$, $k=1,2$, and the weight $\pi$ of the first component is beta, with $\pi \sim \Be(\bar a_\pi, \bar b_\pi)$.

\subsection{Joint Density}
We conclude the exposition of the model by giving the joint density of all parameters, latent variables and observations, making explicit all conditional independence relationships.
We refer to the model as the all-duration flexible SCD model (FSCD).
Let $\regime$, $\state$ and $\obs$ be the flat vectors of all indicators, states and durations, respectively.
Then the joint density is
\begin{align}
\begin{split}
	p(\phi,\sigma,\delta,\tau,\beta,&\xi,\lambda,\pi,\regime,\state,\obs) = \\
	&p(\phi,\sigma) p(\delta \vbar \tau) p(\tau) p(\beta) p(\xi) p(\lambda) p(\pi) \\
	&\prod_{d=1}^D \prod_{i=1}^{n_d}  
		p(\ydi \vbar \sdi, \xdi, \beta, \lambda, \pi) 
		p(\xdi \vbar \xdim, \tdim, t_{d,i-2}, \phi, \sigma, \delta)
		p(\sdi \vbar \sdim, \xi).
\end{split}
\end{align}
Here, the densities for the initial values $\sdf$ and $\xdf$ on each day $d$ are understood to be
$
p(\sdf \vbar \regime_{d,0}, \xi) \equiv p(\sdf \vbar \xi)
$
and
$
p(\xdf \vbar \state_{d,-1}, \tdf, t_{d,-1}, \phi, \sigma, \delta) \equiv p(\xdf \vbar \tdf, \sigma, \delta).
$

We also define the regular-duration model (R-FSCD), the special case of our model where all durations are regular, which is suitable for use with data where trades are aggregated into clusters.
The joint density of the regular-duration model is obtained by removing factors related to $\xi$, $\lambda$, $\pi$ and $\regime$ in the above expressions.

\section{Bayesian Inference and Computation}\label{sec:estimation}
Here we describe posterior simulation methods for Bayesian inference in our flexible SCD model.
We do this for the all-duration model; methods for the regular-duration model require only straightforward modifications.
We end this section by presenting some recommended adjustments to the model and simulation methods, when transaction times are truncated to the second.

\subsection{Posterior simulation}
We use Markov chain Monte Carlo (MCMC) to sample the joint posterior distribution of parameters, latent indicators and state variables.
For the all-duration model, there are six Gibbs blocks, updating $(\state,\phi,\sigma)$, $(\state, \beta)$, $(\delta,\tau)$, $\regime$, $\xi$ and $(\lambda,\pi)$.
For the regular-duration model, we only require the first three blocks.
Note that the state sequence $x$ is updated twice, once jointly with $\phi$ and $\sigma$ and once jointly with $\beta$.
We now describe each of the Gibbs blocks in turn.

\subsubsection{\texorpdfstring
{Drawing from $p(\state, \phi, \sigma \vbar \delta, \beta, \regime, \obs)$}{Block 1}}
It is widely known that when there is strong posterior dependence between state variables (here, $x$) and the parameters of their dynamics (here, $\phi$ and $\sigma$), updating both in a single block improves numerical efficiency.
However, in non-linear non-Gaussian state-space models, it is difficult to draw the state sequence as a single block, even when conditioning on the parameters of the state dynamics.
Several methods have been proposed, and many of them have been applied to draw latent volatilities in stochastic volatility models \citep[see for instance][]{KimShepChib98, ChibNardShep02, RichZhan07}.
Here, an additional difficulty is the non-parametric nature of the measurement distribution, which appears to rule out methods based on auxiliary mixture models.
In this paper, we use the HESSIAN method of \cite{McCa12}, a procedure to draw the state sequence as a single block that does not require data augmentation.
\cite{McCa12} shows how to adapt it to jointly draw the state sequence and associated parameters, which we do here.

Recall that we specify a prior for $\theta \equiv (\log \phi, \log \sigma)$.
Let $\theta_{-}$ be the vector of all parameters except $\theta$.
Our joint proposal $(\state^*, \theta^*)$ consists of a proposal $\theta^*$ drawn from a proposal density $q(\theta \vbar \obs)$ followed by a conditional proposal $\state^*\vbar\theta^*$ drawn from a proposal density $q(\state \vbar \theta, \theta_{-}, \regime, \obs)$.
We accept the pair $(\state^*, \theta^*)$ with probability 
\[
	\min
	\left \{
	1,\,
	\frac
	{
	p(\obs \vbar \theta^*, \theta_{-} ,\regime, \state^*)
	p(\state^* \vbar \theta^*, \theta_{-})
	p(\theta^*)}
	{
	p(\obs \vbar \theta, \theta_{-}, \regime, \state) 
	p(\state \vbar \theta, \theta_{-}) 
	p(\theta)
	}
	\times
	\frac
	{
	q(\theta \vbar \obs)
	q(\state \vbar \theta, \theta_{-},\regime, \obs)
	}
	{
	q(\theta^* \vbar \obs)
	q(\state^* \vbar \theta^*, \theta_{-},\regime, \obs)
	}
	\right \}.
\]
The proposal density $q(\theta \vbar \obs)$ is a multivariate Student's $t$ density  
$t_\nu(\hat \theta, \hat \Sigma_\theta)$ with $\nu=15$ degrees of freedom, where $\hat \theta$ and $\hat \Sigma_\theta$ are approximations of the mean and covariance of the marginal posterior distribution of $\theta$, computed during a burn-in period.
During the burn-in period, we use the adaptive random walk Metropolis approach described in \cite{Viho12}.
At the end of the burn-in period, we compute $\hat \theta$ and $\hat \Sigma_\theta$ as the sample mean and covariance for the second half of the burn-in period.

We draw $\state \vbar \theta^*,\theta_{-},\regime, \obs$ using the HESSIAN method.
In general terms, and suppressing notation for any parameters there may be, this gives a close approximation $q(\state \vbar \obs)$ of the conditional density $p(\state \vbar \obs)$ of the state sequence $\state$ given the observed sequence $\obs$, for state-space models with univariate states in which $p(\obs \vbar \state) = \prod_{i=1}^n p(\obs_i \vbar \state_i)$ and $\state \sim \mathcal N(\bar \Omega^{-1}\bar c, \bar \Omega^{-1})$, with $\bar \Omega$ tridiagonal.
Tridiagonality of $\bar \Omega$ corresponds to $\state$ being Markov but not necessarily homogenous.
The method is generic, as the only model-specific code required consists of a routine to evaluate $\log p(\obs_i \vbar \state_i)$, and its first five derivatives with respect to $\state_i$, at a given point.
For our stochastic duration model, we compute exact values of these derivatives without deriving analytic expressions for them; instead, we exploit automatic routines to combine evaluations of derivatives of primitive functions using Fa\'a di Bruno's rule, which is much easier.
Details are provided in Appendix~\ref{app:derivatives}.

\subsubsection{\texorpdfstring
{Drawing from $p(\state, \beta \vbar \phi, \sigma, \delta, \regime, \obs)$}{Block 2}}
In computational experiments not reported here, we discovered that numerical efficiency is improved by drawing $(\state, \beta)$ as a block, compared to drawing $\beta$ alone, even accounting for the additional computation required to draw $\state$ again in a second block.
Recall that $\state$ and $\beta$ govern the scale and shape, respectively, of the conditional distribution of regular durations.

We update $(\state, \beta)$ in almost the same way as we update $(\state, \theta)$.
For the purposes of drawing proposals, we use the logistic transformation $\vartheta(\beta) = (\log(\beta_1/\beta_J),\dots,\log(\beta_{J-1}/\beta_J))$, which maps the $J-1$ dimensional simplex to $\mathbb{R}^{J-1}$.
The absence of positivity and adding-up constraints is convenient, as is the fact that the posterior distribution of $\vartheta$ is more nearly Gaussian than that of $\beta$, especially when some elements of $\beta$ have high posterior density close to zero.

Let $\beta_{-}$ be the vector of all parameters except $\beta$.
The joint proposal $(\state^*, \beta^*)$ consists of a proposal $\vartheta^*$ drawn from a proposal density $q_\vartheta(\vartheta \vbar y)$, which is transformed back to $\beta^*$ using the inverse transformation $\beta(\vartheta) = (1+\sum_{j=1}^{J-1} \exp(\vartheta_j))^{-1} (\exp(\vartheta_1), \ldots, \exp(\vartheta_{J-1}), 1)$, followed by a conditional proposal $\state^*|\beta^*$ drawn from a proposal density $q(\state \vbar \beta^*, \beta_{-}, \regime, y)$.
We accept $(\state^*,\beta^*)$ with probability
\[
	\min
	\left \{
	1,\,
	\frac
	{
	p(\obs \vbar \beta^*, \beta_{-} ,\regime, \state^*)
	p(\state^* \vbar \beta^*, \beta_{-})
	p(\beta^*)}
	{
	p(\obs \vbar \beta, \beta_{-}, \regime, \state) 
	p(\state \vbar \beta, \beta_{-}) 
	p(\beta)
	}
	\times
	\frac
	{
	q_\beta(\beta \vbar \obs)
	q(\state \vbar \beta, \beta_{-}, \regime, \obs)
	}
	{
	q_\beta(\beta^* \vbar \obs)
	q(\state^* \vbar \beta^*, \beta_{-}, \regime, \obs)
	}
	\right \},
\]
where the implied proposal density $q_\beta(\beta \vbar \obs)$ for $\beta$ is given by $q_\beta(\beta \vbar \obs) = q_\vartheta(\vartheta(\beta) \vbar \obs) / \prod_{i=1}^J \beta_j$ and the proposal density $q_\vartheta(\vartheta \vbar \obs)$ for the transformed parameter $\vartheta$ is a $(J-1)$-variate Student's $t$ with $\nu=15$ degrees of freedom, and mean and variance parameters determined during a burn-in period in exactly the same way as the proposal density $q(\theta \vbar \obs)$, described previously.

Since the $\beta_j$ can be interpreted as mixture weights, it might seem that data augmentation might be useful here.
However, $\beta$ does not only give the mixture weights defining the non-parametric distribution $G(\cdot)$.
It also determines the hazard parameter of the exponential distribution $F(\cdot)$, through the unit-mean normalization of the conditional duration density for regular durations.

\subsubsection{\texorpdfstring
{Drawing from $p(\delta, \tau \vbar \phi, \sigma, \state, \obs)$}{Block 3}}
We update $(\delta,\tau)$ using two sub-blocks.
We first draw $\tau$ from its conditional posterior distribution: $\dbar s \tau \vbar \delta \sim \chi^2(\dbar \nu),$ where $\dbar s = \bar s + \delta' \Delta' \Delta \delta$ and $ \dbar \nu = \bar \nu + L - 1$.
We then draw $\delta$ from its conditional posterior distribution: $\delta \sim N(\dbar \mu, \dbar H)$,
where the posterior precision is $\dbar H = \bar H + W'W$, the posterior mean is $\dbar \mu = \dbar H^{-1} \dbar c$ and the posterior covector is $\dbar c = \bar c + W'\tilde{v}$.
The vector $\tilde{v}$ and matrix $W$ come from writing the state equation \eqref{eq:dynamic-OU} as $\tilde{v} \sim \mathcal N(W\delta,I_N)$ where $v$ and $W$ are organized in blocks
\[
  \begin{bmatrix} \tilde{v}_1 \\ \vdots \\ \tilde{v}_D \end{bmatrix},
  \quad
  \begin{bmatrix} W_{11} & \hdots & W_{1L} \\ \vdots & \ddots & \vdots \\ W_{D1} & \hdots & W_{DL} \end{bmatrix},
\]
with, for $d = 1,\ldots,D$ and $l = 1,\ldots,L$,
\[
  \tilde{v}_d = \begin{bmatrix}
    x_{d,1} / \sigma \\
    (x_{d,2} - \exp(-\phi y_{d,1}) x_{d,1}) / \sqrt{\sigma^2(1 - \exp(-2\phi y_{d,1}))} \\
    \vdots \\
    (x_{d,n_d} - \exp(-\phi y_{d,n_d-1}) x_{d,n_d-1}) / \sqrt{\sigma^2(1 - \exp(-2\phi y_{d,n_d-1}))}
   \end{bmatrix}
\]
and
\[
  W_{dl} = \begin{bmatrix}
    B_l(t_{d,1}) / \sigma \\
    (B_l(t_{d,2}) - \exp(-\phi y_{d,1}) B_l(t_{d,1})) / \sqrt{\sigma^2(1 - \exp(-2\phi y_{d,1}))} \\
    \vdots \\
    (B_l(t_{d,n_d}) - \exp(-\phi y_{d,n_d-1}) B_l(t_{d,n_d-1})) / \sqrt{\sigma^2(1 - \exp(-2\phi y_{d,n_d-1}))}
  \end{bmatrix}.
\]

\subsubsection{\texorpdfstring
{Drawing from $p(\regime \vbar \beta, \xi, \lambda, \pi, \state, \obs)$}{Block 4}}
Latent indicators are updated via a single-move, or one-at-a-time, sampler.
As \cite{Fruh06} points out, single-move sampling is faster than drawing states as a block, as no filtering is required.
Since there is little posterior autocorrelation (most probabilities are close to zero or one, regardless of past and future values) there is little loss of numerical precision.
The relative conditional probabilities of drawing $\sdi=0$ and $\sdi=1$, given $\regime_{-(d,i)}$, the rest of the indicators, are given by
\begin{equation}
	\Pr(\sdi=j \vbar \regime_{-(d,i)}, \beta, \xi, \lambda, \pi, \state, \obs)
	\propto p(\ydi \vbar \sdi =j, \beta, \lambda, \pi, \xdi)\xi_{\sdim,j}\xi_{j,\sdip},
\end{equation}
for $d=1,\dots,D$ and $i=1,\ldots,n_d$.
Here, the first and the last conditional transition probabilities are understood to be $\xi_{j,\regime_{d,0}} \equiv \xi_j$ and $\xi_{j,\regime_{d,n_d+1}} \equiv 1$.

\subsubsection{\texorpdfstring
{Drawing from $p(\xi \vbar \regime)$}{Block 5}}
Given the latent indicators, we update the transition probabilities using a Metropolis-Hasting step.
The prior for $(\xi_{00},\xi_{11})$ ($\xi_{00}$ and $\xi_{11}$ are independent betas) is nearly conditionally conjugate, but since the first indicator of each day comes from the marginal distribution, not exactly so.
The target density can be written as
\begin{equation}
    p(\xi \vbar \regime) \propto
    \left[
    \prod_{d=1}^{D} 
    \frac{1-\xi_{11}^{1-\sdf}-\xi_{00}^{\sdf}}{2-\xi_{00}-\xi_{11}}
    \right]
    \xi_{00}^{N_{00} + \bar a_0 - 1} (1-\xi_{00})^{N_{01} + \bar b_0 - 1}
    \xi_{11}^{N_{11} + \bar a_1 - 1} (1-\xi_{11})^{N_{10} + \bar b_1 - 1},
\end{equation}
where $N_{lk} = \sum_{d,i} \mathbf 1\{\sdi =l, \sdip = k\}$ is the number of transitions from $l$ to $k$ over all $D$ days.
We draw a proposal using two independent beta distributions, $\xi_{00}^* \vbar \regime \sim \Be(N_{00} + \bar a_0, N_{01} + \bar b_0)$ and $\xi_{11}^* \vbar \regime \sim \Be(N_{11} + \bar a_1, N_{10} + \bar b_1)$.
This would be an exact draw from the conditional posterior if we were conditioning on the first indicator in each day.
We correct for the approximation by accepting the proposal with probability 
\[
	\min 
	\left\{ 
	1,
	\prod_{d=1}^D
	\left(\frac{1-\xi^*_{11}}{1-\xi_{11}}\right)^{1-\sdf}
	\left(\frac{1-\xi^*_{00}}{1-\xi_{00}}\right)^{\sdf}
	\left(\frac{2-\xi_{00}-\xi_{11}}{2-\xi^*_{00}-\xi^*_{11}}\right)
	\right\}.
\]

\subsubsection{\texorpdfstring
{Drawing from $p(\lambda, \pi \vbar \regime, \obs)$}{Block 6}}
Recall that cluster durations have a mixture distribution governed by a vector $(\pi,1-\pi)$ of mixture weights and a vector $\lambda = (\lambda_1, \lambda_2)$ of exponential rate parameters.
To update $(\lambda, \pi)$, we first draw component indicators $z_{d,i} \in \{1,2\}$ for each $(d,i)$ for which the duration $\ydi$ is a cluster duration; that is, for which $\sdi = 0$.
The indicators $z_{d,i}$ are conditionally independent, with probability mass function given, up to a multiplicative factor, by
\begin{equation}
	\Pr[z_{d,i} = z \vbar \pi, \lambda, \kappa, \regime, \obs] 
		\propto
		\begin{cases}
		  \pi \lambda_1 e^{-\lambda_1 \ydi}, & z = 1 \\
          (1-\pi) \lambda_2 e^{-\lambda_2 \ydi}, & z = 2.
        \end{cases}
\end{equation}
We then draw $(\lambda,\pi)$ from its conditional distribution given all the $z_{di}$ and $y_{di}$ where $\sdi=0$.
The priors for $\lambda_1$, $\lambda_2$ and $\pi$ are conditionally conjugate, and the conditional posterior distributions are
$
	\lambda_k \vbar z, \regime, \obs 
		\sim \Gam(\bar a_{\lambda_k} + \tilde N_k, \bar b_{\lambda_k} + \tilde N_k \tilde y_k),
$
$k=1,2$, and
$
	\pi \vbar z, \regime, \obs \sim \Be(\bar a_\pi + \tilde N_1, \bar b_\pi + \tilde N_2),
$
where $\tilde{N_k} = \sum_{d,i} \mathbf 1\{z_{d,i} = k\}$ and $\tilde y_k = (1/\tilde{N_k})\sum_{d,i} \ydi \mathbf1\{z_{d,i} = k\}$.

\subsection{Adjustment for recording precision}\label{sec:estimation-adjustment}
Discreteness of duration data arising from limited recording precision is usually ignored, and models mostly feature continuous distributions.
For data recorded to millisecond precision or finer, this is innocuous, but for second precision, the truncation error can be a large fraction of many regular durations, leading to estimation bias and making continuous-time models unreliable for hypothesis testing \citep[see.][]{GMMT05,ZhanLiuBai10,SchnKomlAhma10}.
For this reason, we recommend some minor adjustments to the model when using data truncated to the second.
Discreteness is a property of the recording technology and not the underlying process, so we treat the data as a censored continuous-time process.
\cite{BlasHolyToma20} recently proposed an alternative approach based on parametric discrete distributions, with and without zero inflation.

We first replace the conditional density function of regular durations in equation~\eqref{eq:regular} by a probability mass function obtained by integrating over the range of possible values,
 \begin{equation}
	p_1(\ydi \vbar \xdi) = \frac12
	\int_{\ydi-1}^{\ydi+1} 
	e^{-\xdi} p_\epsilon \left(\obs e^{-\xdi}\right)
	d\obs.
\end{equation}
For the probability of $\ydi=0$, the integral is from zero to one.

We then replace the distribution of cluster durations by a Bernoulli distribution with parameter $\zeta$; the values of $0s$ and $1s$ have probabilities $\zeta$ and $(1-\zeta)$, respectively.
We specify a conditionally conjugate beta prior: $\zeta \sim \Be(\bar a_\zeta, \bar b_\zeta)$.

We require some straightforward adjustments to our posterior simulator.
The block updating $(\lambda,\pi)$ is replaced by a block updating $\zeta$.
The conditional posterior distribution for $\zeta$ is $\zeta \vbar \regime, \obs \sim \Be(\bar a_\zeta + N_{00}, N_{11} + \bar b_\zeta)$.
The computation of the derivatives required by the HESSIAN method needs straightforward adjustment.

\section{Results}\label{sec:results}
Here, we first report results from an artificial data experiment meant to test for the correctness of our posterior simulators.
We then illustrate the use of the flexible SCD model with an empirical application, using transaction data for two equities traded on the Toronto Stock Exchange.

\subsection{Getting it right (GIR)}
When posterior simulation methods are based on incorrect analysis or when there are coding errors in their implementation, we cannot rely on the results they produce.
The tests for program correctness described here are similar to those described in \cite{Gewe04} and the title of this section comes from the title of that paper.

\begin{table}[t]
\begin{center}
\begin{threeparttable}
	\caption
	{\small
	Prior hyper-parameter values used in the Getting it right experiment (GIR) and the empirical application with TSX data.
	}
	\label{tab:hyper-parameters}
	\begin{tabular}{l l l}
		\toprule
		Hyper-parameters
		& GIR & TSX \\
		\midrule
		$\bar \theta$ 
		& $(-3.5, -1.5)$ & $(-4.5, -1.5)$ \\
		$(\bar \Sigma_{11};\bar \Sigma_{22}; \bar \Sigma_{12})$ 
		& $(0.01; 0.01; 0.00)$ & $(0.25; 0.05; -0.05)$ \\
		$(\bar \delta, \bar h)$  				
		& $(1.0, 500)$ & $(2.5, 2.0)$ \\
		$(\bar s, \bar \nu)$
		& $(5.0, 1000)$ & $(1.0, 100)$ \\
		$(\bar M; \bar \beta)$					   
		& $(500; 0.5, 0.3, 0.2)$ & $(10J; 1/J,\dots,1/J)$ \\
		$(\bar a_0, \bar b_0)$					 
		& $(200, 300)$ & $(3, 2)$ \\
		$(\bar a_1, \bar b_1)$		
		& $(400, 100)$ & $(2, 3)$ \\
		$(\bar a_{\lambda_1}, \bar b_{\lambda_1})$ 
		& $(500, 5)$ & --- \\
		$(\bar a_{\lambda_2}, \bar b_{\lambda_2})$ 
		& $(500, 10)$ & ---	\\
		$(\bar a_\pi, \bar b_\pi)$
		& $(250, 250)$ & --- \\
		$(\bar a_\zeta, \bar b_\zeta)$
		& $(475, 25)$ & $(15, 2)$ \\
		\bottomrule
	\end{tabular}
\end{threeparttable}
\end{center}
\end{table}

The idea of the exercise is to simulate a Markov chain whose stationary distribution is the {\em joint} distribution of parameters, latent state variables {\em and data}, making use of the same simulation methods that will be used later for posterior simulation, together with an additional Gibbs block to draw data from their conditional distribution given parameters and state variables.
If the posterior simulation methods are correct in concept and implementation, the marginal distribution of the parameter vector, with respect to this stationary distribution, is identical to its (known) prior distribution.
This is a strong condition with many easily testable implications.

We will need to simplify part of our model in order to proceed.
The problem is that we record the values of the underlying continuous-time OU process $\state(t)$ only at the times that trades occur.
Redrawing duration data changes the trading times, which requires conditioning on the entire path of the continuous-time process, which is impractical.
For the GIR simulations only, we modify the latent state process described in \eqref{eq:dynamic-OU}, replacing the OU process with a homogenous autoregressive process where the transition distribution from $\state_i$ to $\state_{i+1}$ depends only on $\state_i$ and not the duration $\obs_i$.
We parameterize the process in a way that resembles the sampled OU process, giving
\begin{equation}\label{eq:dynamic-gir}
	\state_{i+1} \vbar \state_i, t_i, t_{i-1}  \sim \mathcal N 
		\left( m(t_i) + e^{-\phi}(\state_i - m(t_{i-1})), \sigma^2 (1-e^{-2\phi}) \right).
\end{equation}

The additional Gibbs block updating the data $\obs$ from its conditional distribution given parameters and latent variables is described in Appendix~\ref{app:drawing-y}.
Sampling from the posterior distribution requires some minor modifications.
Since we draw a new sample of artificial data at each iteration, the adaptive schemes implemented to approximate the mean and the covariance of $p(\theta \vbar \obs)$ and $p(\beta \vbar \obs)$ during the burn-in period do not work well.
Instead, we use independence Metropolis-Hastings updates, where the proposal distribution has the same mean and covariance as the parameter vector in question.\footnote{
	We approximate the mean and covariance matrix of $p(\vartheta)$ for $\vartheta(\beta) = (\log(\beta_1/\beta_J),\dots,\log(\beta_{J-1}/\beta_J))$by
	\begin{align*}
		\hat \vartheta_{j} &= \psi(\bar M \bar \beta_j) - \psi(\bar M \bar \beta_J)
		\quad j=1,\dots,J-1\\
		\hat \Sigma_{jj} &= \psi'(\bar M \bar \beta_j) + \psi'(\bar M \bar \beta_J)
		\quad j=1,\dots,J-1 \\
		\hat \Sigma_{jk} &= \psi'(\bar M \bar \beta_J) \quad j \neq k
	\end{align*}
	where $\psi$ is the digamma function and $\psi'$ the trigamma functions.
	Those relations are the solution of the approximation of a Dirichlet distribution by a logistic normal distribution minimizing their Kullback-Leibler divergence.
}
We set the number of components of the normalized distribution to $J=3$ and use a B-spline function defined on two knots, $\topen$ and $\tclose$, giving a diurnal pattern that is an expansion with $L=4$ cubic polynomials.
To avoid trades in simulations occurring after $\tclose$, where the diurnal pattern is undefined, we choose the sample size $n$ (the number of durations, not the size of the simulation sample) and prior distributions such that the probability that the last transaction of the day occurs after $\tclose$ is extremely small.
We fix $D=1$ and $t_0=\topen$, choose a sample size of $n=50$ observations, and set the length of the trading session to $600$ seconds.
Values of the prior hyper-parameters are shown in Table~\ref{tab:hyper-parameters}.
The tighter prior distribution, and much smaller number of observations, compared with our empirical application, ensure high numerical precision with a moderate amount of computation.

\begin{table}[t]
\begin{center}
\begin{threeparttable}
	\caption
	{\small
	Difference between prior means and simulation sample means in the Getting it right experiment.
	}
	\label{tab:results-getting-it-right}
	\begin{tabular}{c rrr rrr }
		\toprule
		\multicolumn{1}{c}{}			 	&
	   	\multicolumn{3}{c}{Continuous} 	 	&
	    \multicolumn{3}{c}{Truncated} 	 	\\
		$\vartheta$ 					 	& 
		$\E[\vartheta] - \bar \vartheta$  	&
		$\hat \sigma_{\mathrm{nse}}$		&	 
		$t$-stat       				 		&
		$\E[\vartheta] - \bar \vartheta$ 	&
		$\hat \sigma_{\mathrm{nse}}$		&	 
		$t$-stat         				 	\\
	    \cmidrule(lr){1-1}
		\cmidrule(lr){2-4}
		\cmidrule(lr){5-7}
		\input{results-getting-it-right.tex}
		\bottomrule
	\end{tabular}
	{\footnotesize
	\textit{Note.}
	$\theta = (\log \phi, \log \sigma)$
	}
\end{threeparttable}
\end{center}
\end{table}

We verify the correctness of both the continuous-time model and the discrete-time version that accommodates the truncation of trading times to the second.
We generate a sample of size $10^6$ in each case.
Table~\ref{tab:results-getting-it-right} shows the results, for the continuous-time model on the left and the discrete-time model on the right.
In each case, the first column gives, for selected parameters $\vartheta$, the difference between the prior mean $E[\vartheta]$ and the simulation sample mean $\bar\vartheta$; the second, the numerical standard error (i.e.\ the simulation standard deviation quantifying error in finite simulation) of the sample mean; and the third, the $t$-statistic for the test of the hypothesis that the mean of the parameter, with respect to the stationary distribution of the Markov chain, equals the (known) prior mean.
Numerical standard errors are computed using the overlapping batch means method \citep[see.][]{FlegJone10}.
Rather than reporting results for all elements of $\delta$, we report results for their mean $v'\delta$ and the smoothing parameter $\tau$.
Sample means are close to the true prior means, relative to the numerical standard error.
Each of the hypotheses is a necessary condition for the correctness of our simulation methods.
The results fail to cast doubt on this correctness: one hypothesis (out of 22) is rejected at the 5\% level and none at the 1\% level.

\subsection{Empirical Application}
We demonstrate our FSCD model and posterior simulation methods using the transaction data described in Section~\ref{sec:data}.
The data, from March 2014, are for two equities traded on the Toronto Stock Exchange: the Royal Bank of Canada (RY) and the Potash Corporation (POT).
We analyze the full samples directly using the all-duration model.
To compare our probabilistic approach to classifying durations as regular or cluster durations with deterministic aggregation rules, we also construct subsamples of the durations classified as regular by the GW aggregation rule.
We will refer to these subsamples as GW-filtered subsamples, and analyze them using the regular-duration model (R-FSCD).
We remind the reader that the same-second aggregation rule of common practice is cruder than the GW aggregation rule; we would expect distortions arising from the same-second aggregation rule to be even more serious than those reported here.

\subsubsection{Model specification and prior distributions}
For both specifications (all-duration and regular-duration) we report results for four models, each with a fixed value of $J$, the number of terms in the normalized conditional density; those values are $J=2,3,4$ and $5$.
Rather than estimate $J$, which is difficult in models with two different state variables, we present results for each value of $J$ and compare them.
In all models, the diurnal pattern is specified as a B-spline function defined on knots set on each half-hour, giving an expansion with $L=16$ piecewise polynomials.\footnote{
	Trade durations are often standardized using a cubic spline specification with knots set at each hour, with extra knots in the first and last half hour to better capture rapid changes of trading intensity at the beginning and end of the trading day.
}
Values of the prior hyper-parameters are shown in the third column of Table~\ref{tab:hyper-parameters}.
We select a fairly diffuse prior distribution for the log-transformed parameters of the 
latent intensity state $\xdi$, the mean of the coefficients of the B-spline function $\mean(t)$ describing diurnal patterns and the transition probabilities of the latent indicator $\sdi$.
The values of $(\bar s, \bar \nu)$ give a more diffuse prior distribution for the smoothing parameter $\tau$ than the one suggested by \cite{LangBrez04}.
For reasons given in the introduction, we center the prior distribution for the flexible conditional duration density around the exponential distribution.
We set the concentration parameter for moderate shrinkage towards the exponential distribution: most of posterior variances of the $\beta_j$ coefficients are less than a quarter of the prior variances.
The prior distribution for $\zeta$ favours cluster durations of $0s$ over $1s$, without being too informative.

\subsubsection{Analysis of trade durations}
\begin{sidewaystable}
\begin{center}
\begin{threeparttable} 
    \caption
    {\small
    Posterior mean and standard deviation of various parameters, for the full RY sample and the all-duration model; and for the GW-filtered RY subsample and the regular-duration model.
    }
    \label{tab:results-ry}
    \begin{tabular}{l *3c *3c *3c *3c}
    	\toprule
    	\multicolumn{13}{l}{{\bf Panel A: All-duration models}} \\
    	\midrule
        & Mean & Std & RNE & Mean & Std & RNE & Mean & Std & RNE & Mean & Std & RNE \\
        \cmidrule(lr){1-1}
		\cmidrule(lr){2-4}
		\cmidrule(lr){5-7}
		\cmidrule(lr){8-10}
		\cmidrule(lr){11-13}
        \input{results-param-all-ry.tex}
    \end{tabular}
    \begin{tabular}{l *3c *3c *3c *3c}
    	\toprule
    	\multicolumn{13}{l}{{\bf Panel B: Regular-duration models}} \\
    	\midrule
		& Mean & Std & RNE & Mean & Std & RNE & Mean & Std & RNE & Mean & Std & RNE \\
        \cmidrule(lr){1-1}
		\cmidrule(lr){2-4}
		\cmidrule(lr){5-7}
		\cmidrule(lr){8-10}
		\cmidrule(lr){11-13}
        \input{results-param-reg-ry.tex}
        \bottomrule
    \end{tabular}
    {\footnotesize
    $\dagger$: Rescale $\times 100$. \\
    \textit{Note.}
    The table gives the posterior mean and standard deviation, and the relative numerical efficiency (RNE) for the posterior mean, based on 50,000 posterior draws recorded after a burn-in period of 15,000 draws.
   	The four groups of these three columns are for two, three, four and five component normalized duration densities.  
    }
\end{threeparttable} 
\end{center}
\end{sidewaystable}

\begin{sidewaystable}
\begin{center}
\begin{threeparttable} 
    \caption
    {\small
    Posterior mean and standard deviation of various parameters, for the full POT sample and the all-duration model; and for the GW-filtered POT subsample and the regular-duration model.
    }
    \label{tab:results-pot}
    \begin{tabular}{l *3c *3c *3c *3c}
    	\toprule
    	\multicolumn{13}{l}{{\bf Panel A: All-duration models}}\\
    	\midrule
		& Mean & Std & RNE & Mean & Std & RNE & Mean & Std & RNE & Mean & Std & RNE \\
        \cmidrule(lr){1-1}
		\cmidrule(lr){2-4}
		\cmidrule(lr){5-7}
		\cmidrule(lr){8-10}
		\cmidrule(lr){11-13}
        \input{results-param-all-pot.tex}
    \end{tabular}
        \begin{tabular}{l *3c *3c *3c *3c}
    	\toprule
    	\multicolumn{13}{l}{{\bf Panel B: Regular-duration models}}\\
    	\midrule
		& Mean & Std & RNE & Mean & Std & RNE & Mean & Std & RNE & Mean & Std & RNE \\
        \cmidrule(lr){1-1}
		\cmidrule(lr){2-4}
		\cmidrule(lr){5-7}
		\cmidrule(lr){8-10}
		\cmidrule(lr){11-13}
        \input{results-param-reg-pot.tex}
        \bottomrule
    \end{tabular}
    {\footnotesize
    $\dagger$: Rescale $\times 100$. \\
    \textit{Note.}
    The table gives the posterior mean and standard deviation, and the relative numerical efficiency (RNE) for the posterior mean, based on 50,000 posterior draws recorded after a burn-in period of 15,000 draws.
   	The four groups of these three columns are for two, three, four and five component normalized duration densities.  
    }
\end{threeparttable} 
\end{center}
\end{sidewaystable}

Table~\ref{tab:results-ry}~and~\ref{tab:results-pot} show the results for the RY series and the POT series, respectively.
For each parameter, we report the posterior mean and standard deviation, and the relative numerical efficiency (RNE) for the posterior mean.
Defined in \cite{Gewe89}, the relative numerical efficiency is a variance ratio that quantifies the numerical precision of the sample mean of a ergodic process, relative to that of a (hypothetical) iid sample.
RNE times sample size gives the size of an iid sample with the same numerical standard error.
Numerical standard errors are computed using the overlapping batch mean method \citep{FlegJone10}.
The posterior samples consist of 50,000 retained draws recorded after a burn-in period of 15,000 draws.

It is well known that it is difficult to sample efficiently the persistence and variance parameters of latent states in non-Gaussian state-space models.
We obtain a numerical efficiency for $\phi$ and $\sigma$ that is considerably higher than that reported for the analogous parameters using the block sampling method in \cite{StriForbMart06}.\footnote{
	The latent intensity state in their analysis follows a Gaussian AR(1) process with fixed autocorrelation and fixed innovation variance rather than an OU process sampled at irregular intervals.
}
The lowest numerical efficiency we obtain for these parameters is more than ten times the highest numerical efficiency they obtained for analogous parameters, using parametric duration distributions.
Posterior sample means of the mixture weights have numerical efficiency ranging from 0.03 to 0.47 for the all-duration models and from 0.31 to 0.66 for the regular-duration models.
Although numerical efficiencies are lower for the all-duration models, especially for specifications with a more flexible conditional duration density, they are quite good as far as methods for non-linear non-Gaussian state-space models go.

\begin{table}[t]
\begin{center}
\begin{threeparttable}
	\caption
    {\small
    Posterior quantiles and moments of the half-life measured in second $t_{1/2}$ of the latent intensity state process $\xdi$.
    }
    \label{tab:results-half-life}
    \begin{tabular}{l rrrr rrrr}
    	\toprule
    	\multicolumn{1}{c}{}    &
    	\multicolumn{4}{c}{RY}  &
    	\multicolumn{4}{c}{POT} \\
        & Mean & Std & $q_{0.025}$ & $q_{0.975}$ & Mean & Std & $q_{0.025}$ & $q_{0.975}$ \\
        \cmidrule(lr){1-1}
		\cmidrule(lr){2-5}
		\cmidrule(lr){6-9}
        \input{results-half-life.tex}
        \bottomrule
    \end{tabular}
\end{threeparttable} 
\end{center}
\end{table}

Table~\ref{tab:results-half-life} reports posterior quantiles and moments of the half-life $t_{1/2}$, measured in seconds, of the OU process $\state_d(t)$: the length of time it takes for $\state_d(t)$ and $\state_d(t+t_{1/2})$ to have a correlation of 1/2 between them.
This quantity is more easily interpretable than the mean reversion parameter $\phi$.
Persistence of the latent intensity process is fairly high, and more so for RY.
There is a fair degree of posterior uncertainty about $t_{1/2}$, and the posterior distribution is somewhat sensitive to how regular durations are classified and to the number of terms, $J$, in the normalized duration density.
The marginal standard deviation $\sigma$ is estimated more precisely (see Tables~\ref{tab:results-ry}~and~\ref{tab:results-pot}).
Its distribution is somewhat sensitive to how regular durations are classified, but very little to $J$.

\begin{figure}[t]
    \centering
    \begin{subfigure}[b]{0.475\textwidth}
    	\centering
    	\includegraphics[width=\textwidth]
    	{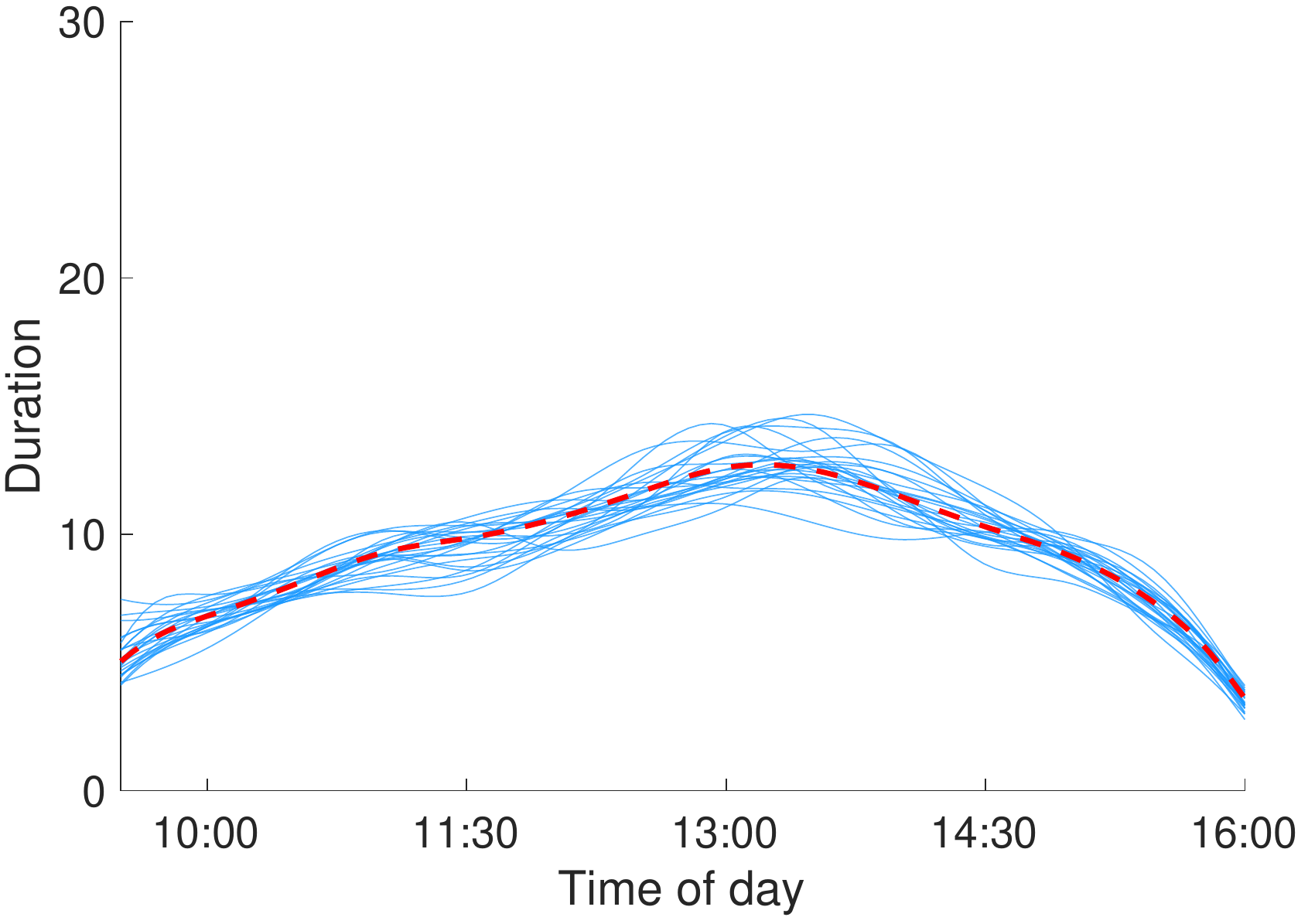}
    \end{subfigure}
    \quad
    \begin{subfigure}[b]{0.475\textwidth}  
    	\centering 
    	\includegraphics[width=\textwidth]
    	{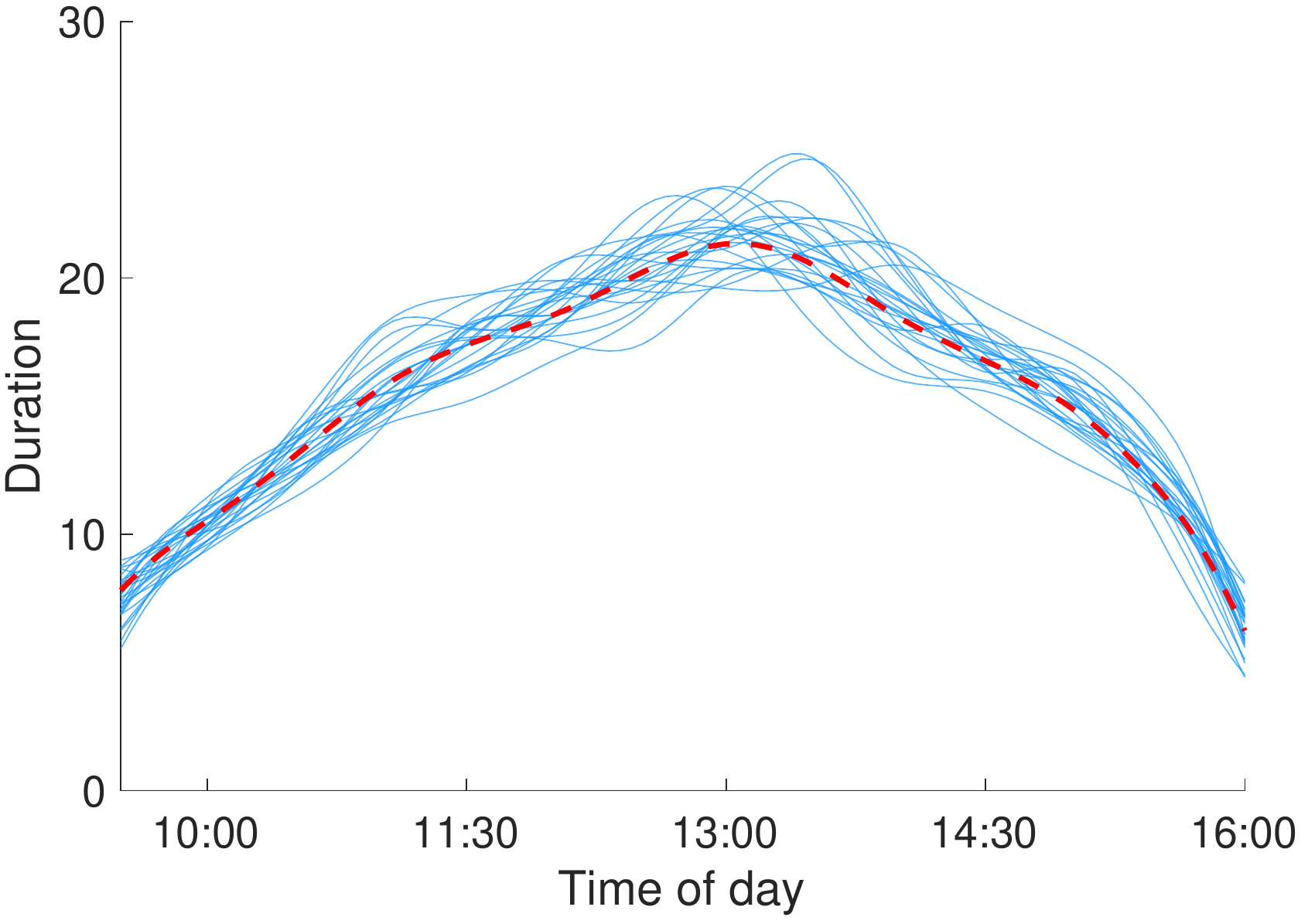}
    \end{subfigure}
	\caption
	{\small
	Diurnal pattern at the posterior mean (dashed line) and for 25 posteriors draws, (solid lines) obtained for the full samples, and the all-duration model with $J=5$.
	The figure on the left is for the RY series and the one on the right, the POT series.
	}
	\label{fig:results-diurnal}
\end{figure}

Figure~\ref{fig:results-diurnal} shows the posterior mean (dashed line) and 25 posterior draws of the diurnal pattern $\mean(t)$, for the full sample and the all-duration model with $J=5$ terms.
The left panel is for the RY series and the right panel for POT.
The diurnal pattern for RY is flatter, indicating less predictable variation in trading intensity.
In both cases, we obtain the usual inverted U-shaped diurnal pattern found in most studies, with more trading intensity near the opening and closing times.
The posterior variation in the diurnal patterns is fairly small compared to the variation in average intensity through the day.
The posterior distribution for $\delta$ is not very sensitive to $J$ or to the choice between deterministic and probabilistic classification.
For this reason, we do not show illustrations similar to Figure~\ref{fig:results-diurnal} for other specifications.

\begin{figure}[t]
    \centering
    \begin{subfigure}[b]{0.475\textwidth}
    	\centering
    	\includegraphics[width=\textwidth]
    	{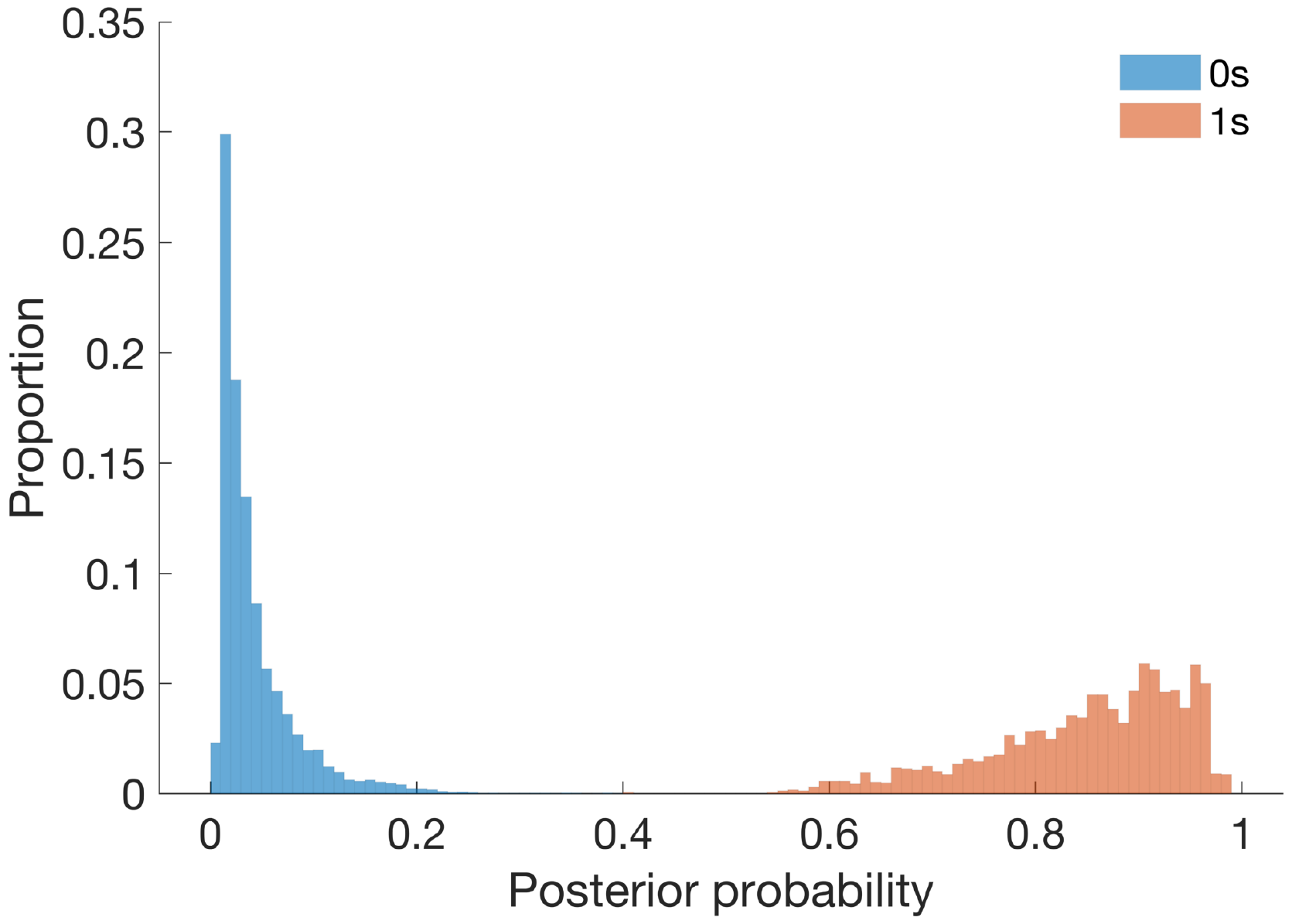}
    \end{subfigure}
    \quad
    \begin{subfigure}[b]{0.475\textwidth}  
    	\centering 
    	\includegraphics[width=\textwidth]
    	{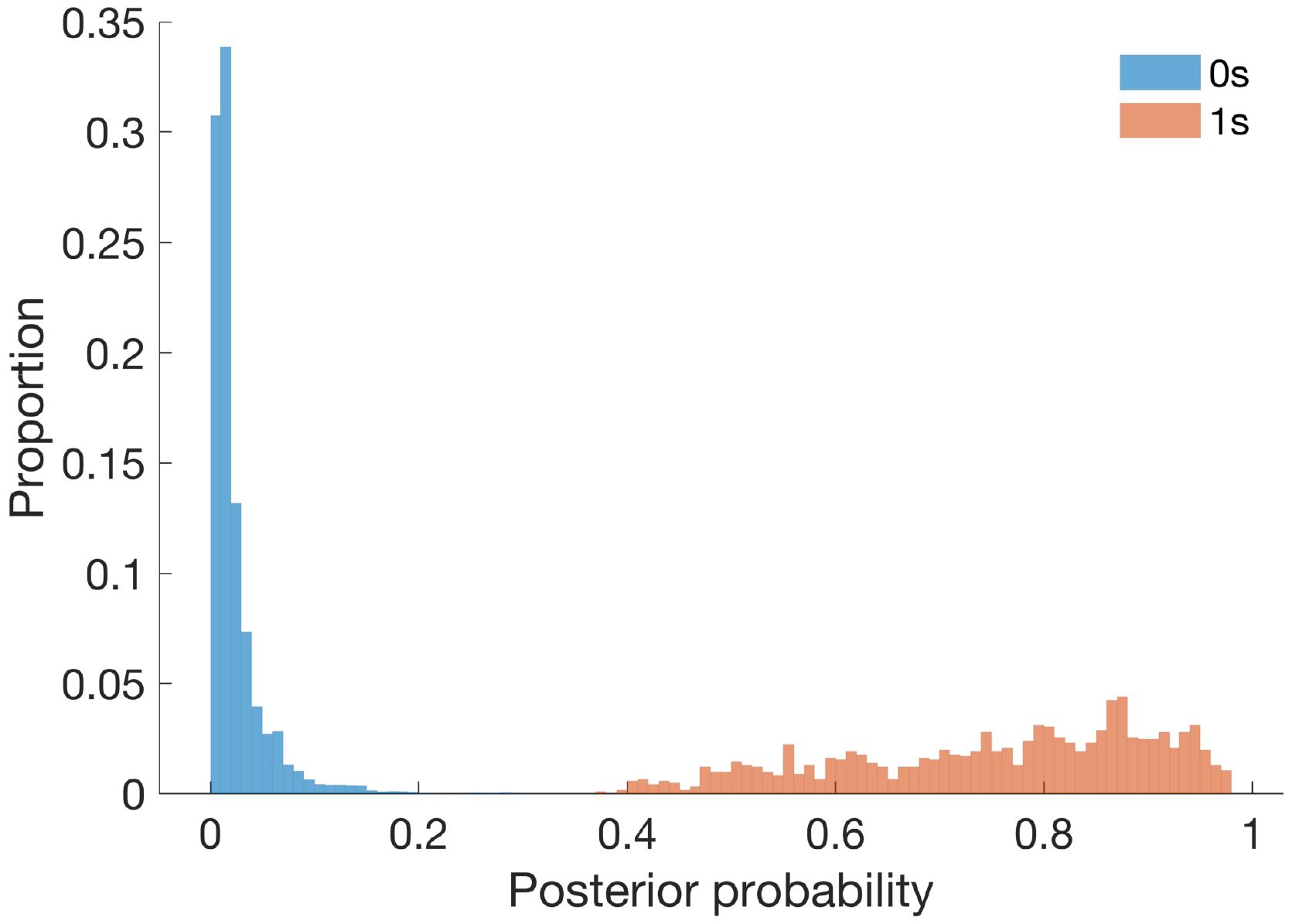}
    \end{subfigure}
    \caption
    {\small
    Histogram of posterior probabilities of being regular, for durations recorded as $0s$ or $1s$, for the full sample and all-duration model with $J=5$.
    The histogram on the left is for the RY series; the one on the right, the POT series.
    The histograms illustrate variation over observations, not posterior uncertainty.
	}
    \label{fig:results-cluster}
\end{figure}

For all-duration models, we record, at each posterior draw, whether each duration recorded as $0s$ or $1s$ is a regular or a cluster duration; classifications vary from draw to draw.
We summarize this in two ways: first, we compute, for each of these durations, the probability that it is regular and illustrate the variation of this probability over durations; second, we describe the posterior distribution of the number of these durations that are regular.
Figure~\ref{fig:results-cluster} shows histograms of classification probabilities for durations recorded as $0s$ and as $1s$ under the all-duration model with $J=5$, for both series.
The horizontal axes give the posterior probability that a duration is regular, and the vertical height of the bar at a given histogram bin gives the proportion of durations recorded as $0s$ or $1s$ whose posterior probability of being regular is within that bin.
Every duration recorded as $0s$ has a posterior probability less than $0.3$ of being a regular duration; for a large majority, it is less than $0.05$.
In contrast, most durations recorded as $1s$ have a posterior probability of more than $0.5$ of being regular.
Histograms for other values of $J$ are similar and we do not report them.

\begin{table}[t]
\begin{center}
\begin{threeparttable}
    \caption
    {\small
    Posterior quantiles and moments for the number of durations recorded as $0s$ and $1s$ classified as regular.
    }
    \label{tab:results-classification}
    \begin{tabular}{c l rrrr rrrr}
    	\toprule
    	\multicolumn{2}{c}{}    &
 		\multicolumn{4}{c}{RY}  &
    	\multicolumn{4}{c}{POT} \\
        &
        & 
        Mean & Std & $q_{0.025}$ & $q_{0.975}$ & 
 		Mean & Std & $q_{0.025}$ & $q_{0.975}$ \\
        \cmidrule(lr){1-1}
        \cmidrule(lr){2-2}
        \cmidrule(lr){3-6}
        \cmidrule(lr){7-10}
        \input{results-classification.tex}
        \bottomrule
    \end{tabular}
\end{threeparttable}
\end{center}
\end{table}

While durations recorded as $0s$ are each quite unlikely to be regular, they are very numerous, and so the probability that many of them are regular is nonetheless very high.
Table~\ref{tab:results-classification} shows posterior quantiles and moments of the number of durations recorded as $0s$ or $1s$ classified as regular.
Results for durations recorded as $0s$ are reported in the first four rows; those for durations recorded as $1s$, in the last four rows.
For comparison, the GW rule classifies 386 durations recorded as $0s$ as regular for the RY series and 187 for the POT series.
The GW rule does not apply to durations recorded as $1s$ and so all 2346 RY and all 1252 POT durations recorded as $1s$ are considered regular.
Our probabilistic approach classifies as regular many more durations recorded as $0s$ and slightly fewer durations recorded as $1s$.
For both series, the posterior standard deviation of their number increases with the number of terms $J$ in the normalized duration density; more terms allows more flexibility in the shape of the normalized conditional density near zero.

To resume: for each duration recorded as $0s$, classifying it as a cluster duration is the right choice, under symmetric loss.
Collectively, these zero/one decisions lead one to severely underestimate the number of these durations that are regular.
With our probabilistic approach, we get a very good idea about how many of these durations are regular, although of course we do not magically discover which ones.
This approach is exactly how we avoid artifacts arising from the spurious aggregation of unrelated, but nearly simultaneous, trades.

\begin{figure}[t]
    \centering
    \begin{subfigure}[b]{0.475\textwidth}
    	\centering
    	\includegraphics[width=\textwidth]
    	{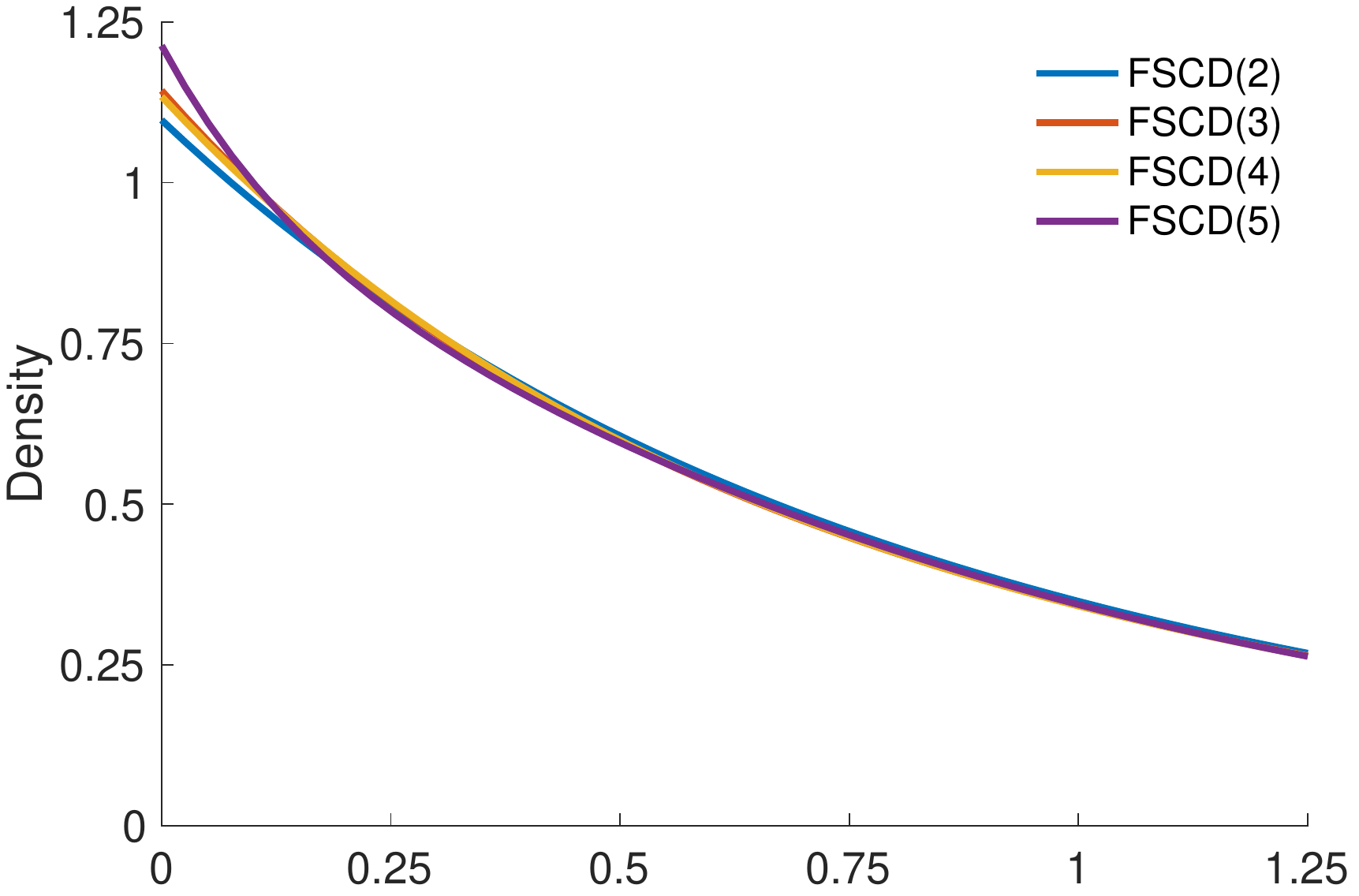}
    \end{subfigure}
    \quad
    \begin{subfigure}[b]{0.475\textwidth}  
    	\centering 
    	\includegraphics[width=\textwidth]
    	{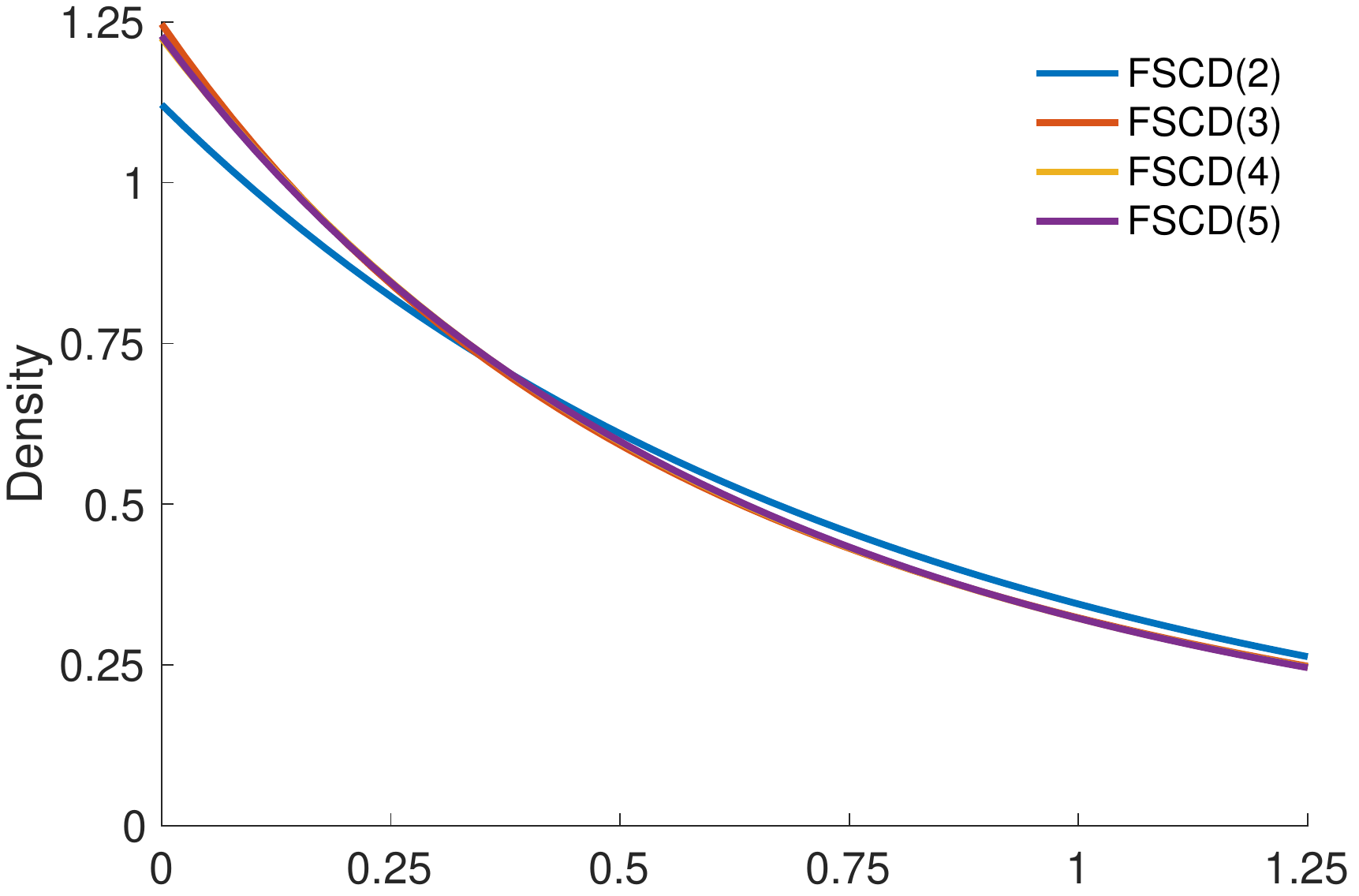}
    \end{subfigure}
	\vskip 1.5\baselineskip
    \begin{subfigure}[b]{0.475\textwidth}
    	\centering
    	\includegraphics[width=\textwidth]
    	{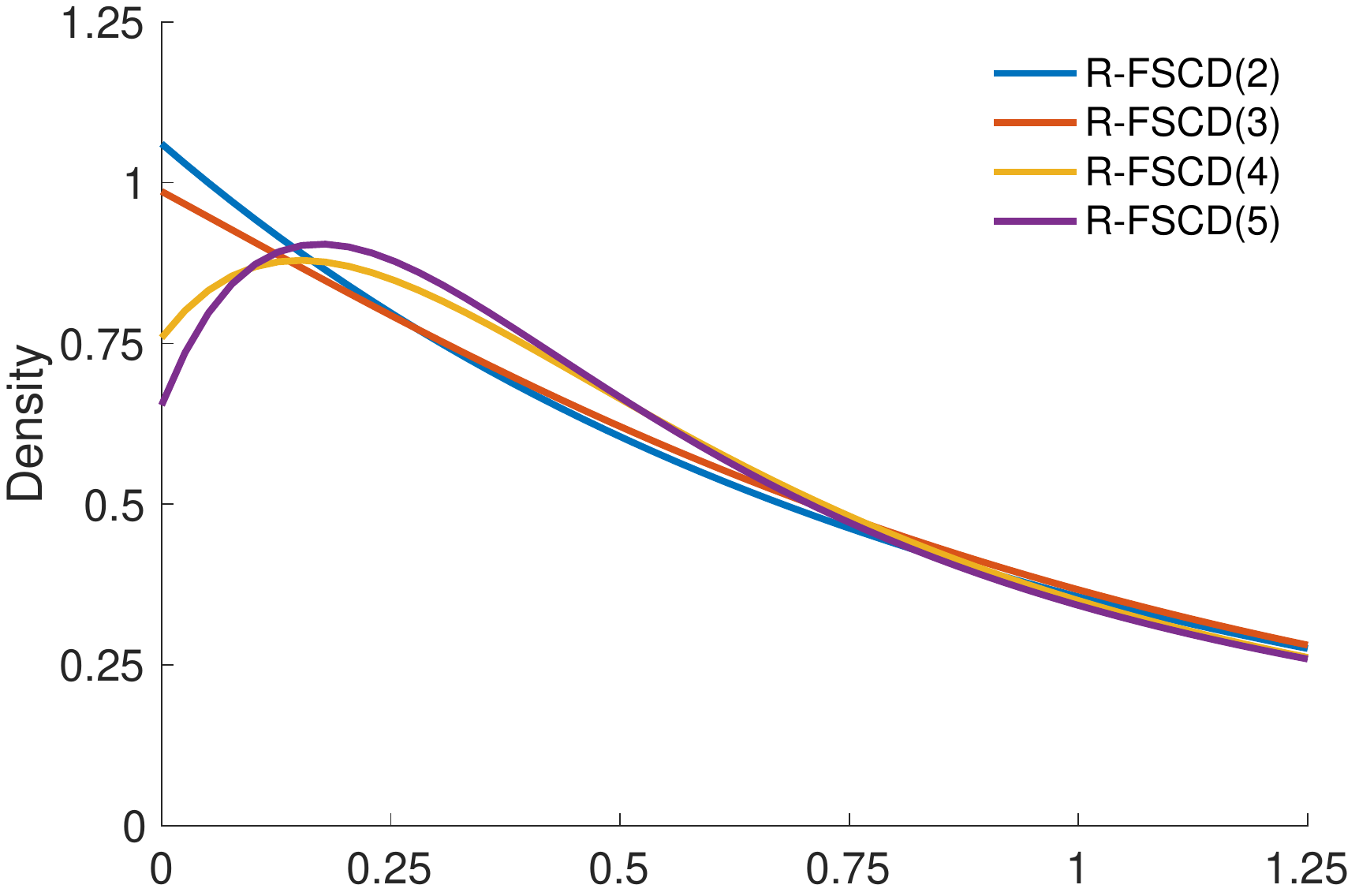}
    \end{subfigure}
    \quad
    \begin{subfigure}[b]{0.475\textwidth}  
    	\centering 
    	\includegraphics[width=\textwidth]
    	{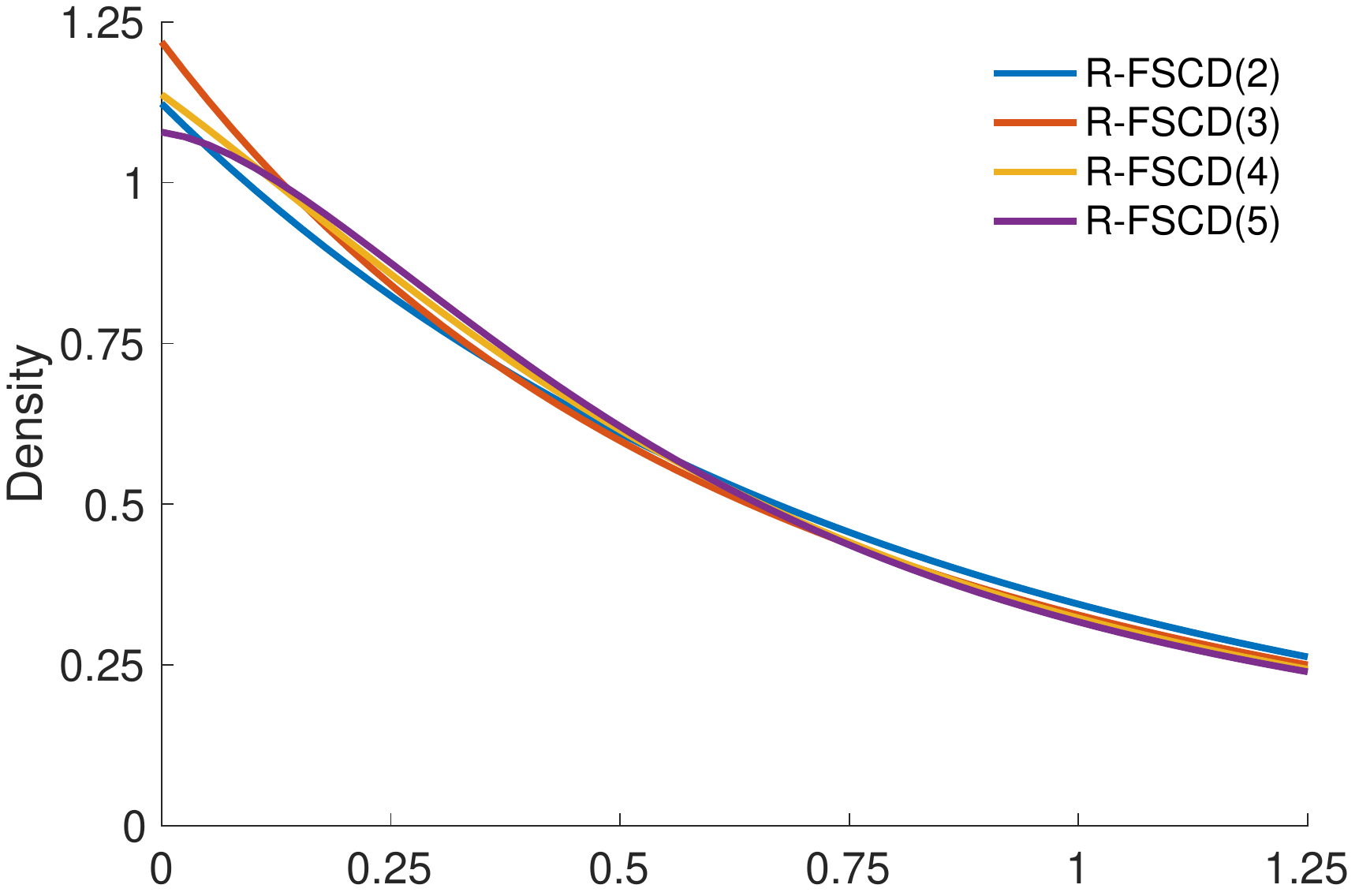}
    \end{subfigure}
    \caption
	{\small
	Normalized density functions at the posterior mean of $\beta$.
	Upper panels are for full samples and all-duration models; lower panels, for GW-filtered subsamples and regular-duration models.
	Panels on the right are for the RY series; panels on the left, for POT.
	}
	\label{fig:results-density}
\end{figure}

\begin{figure}[t]
    \centering
    \begin{subfigure}[b]{0.475\textwidth}
    	\centering
    	\includegraphics[width=\textwidth]
    	{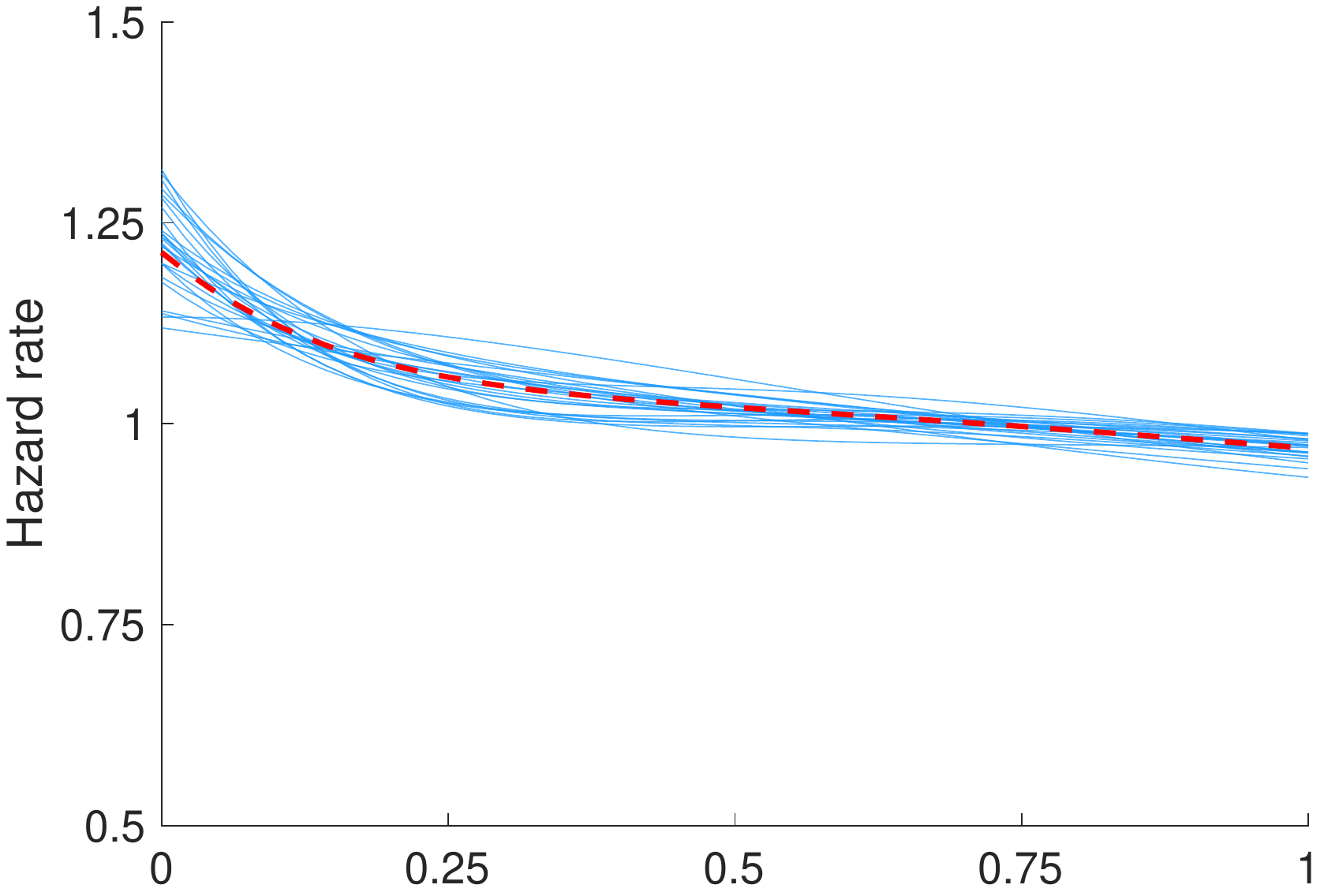}
    \end{subfigure}
    \quad
    \begin{subfigure}[b]{0.475\textwidth}  
    	\centering 
    	\includegraphics[width=\textwidth]
    	{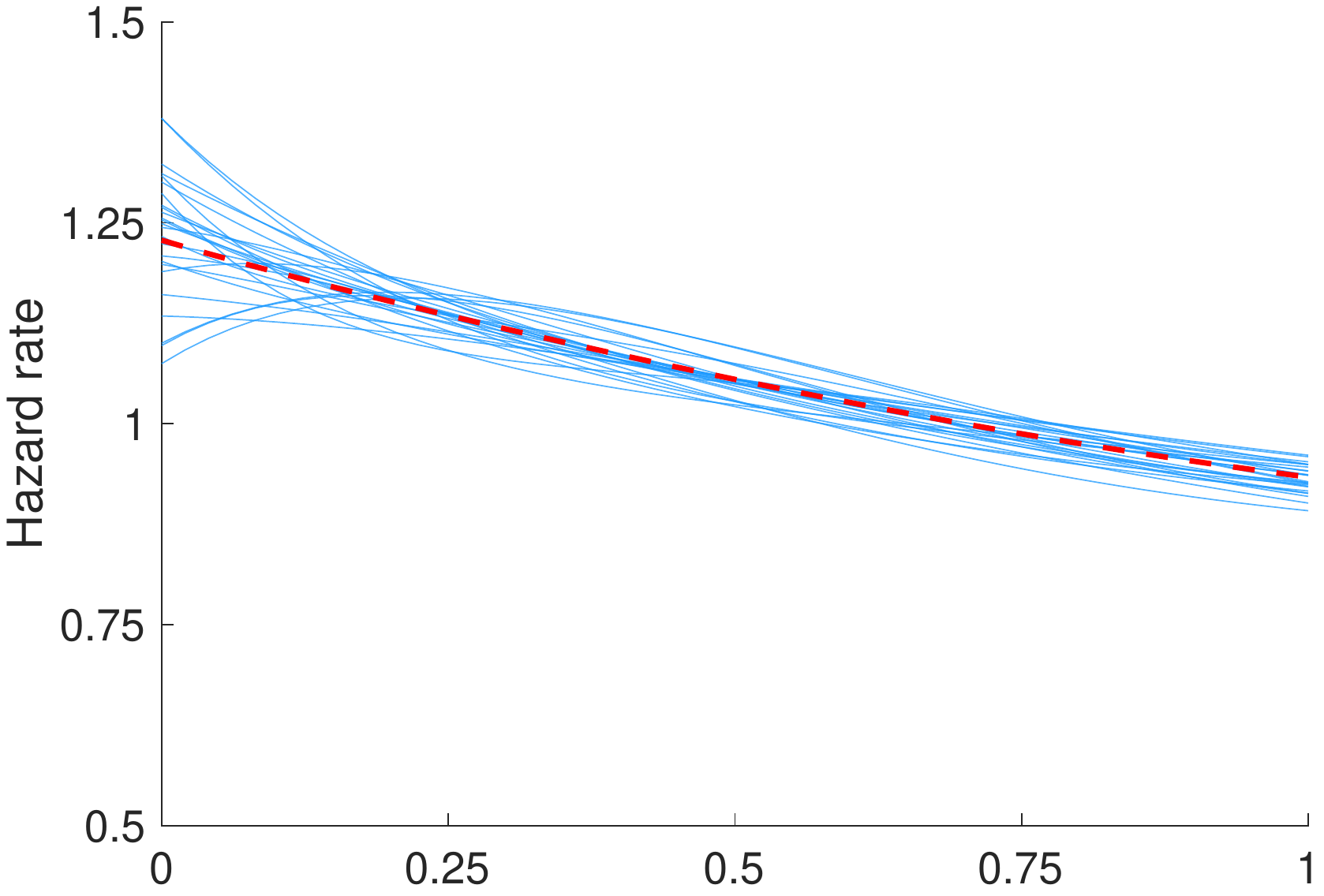}
    \end{subfigure}
    \caption
	{\small
	Hazard functions at the posterior mean (dashed line) and for 25 posteriors draws, (solid lines) obtained for the full samples, and the all-duration model with J = 5. The left figure shows the diurnal pattern for the RY series and the right figure for the POT series
	}
	\label{fig:results-hazard}
\end{figure}

Figure~\ref{fig:results-density} show normalized conditional density functions at the posterior mean of the $\beta$ coefficients.
Upper panels are for the all-duration models and the full sample; lower panels for the regular-duration models and the GW-filtered subsamples; RY on the left and POT on the right.
We see that the conditional density at the posterior mean of $\beta$ varies much more with $J$ at shorter durations than it does at longer durations.
For regular-duration models, the density at zero decreases appreciably in $J$, especially for the RY series.
With extra flexibility, the density increasingly fits the spurious scarcity of durations recorded as $0s$ in the GW-filtered sample; the resulting distortions are seen well away from zero.
For the all-duration models (upper panels), there is less variation of the density at zero, despite the uncertainty about the number of durations recorded as $0s$ that are regular, and little difference between the two series.
For the POT series, the mean densities obtained using $3$, $4$ and $5$ terms are barely distinguishable.
For the all-duration models, the density at zero increases, rather than decreases in $J$, but the variation is not very pronounced, even near zero, where uncertainty about the number of regular durations recorded as $0s$ or $1s$ comes into play.
While we are not able to estimate $J$, results do not change much between $J=4$ and $J=5$, and we suggest that the $J=5$ specification is able to capture well the normalized conditional density.
Figure~\ref{fig:results-hazard} illustrates the posterior distribution of the hazard function for the normalized conditional duration density, for the full samples and the all-duration model with $J=5$.
The dashed line shows the hazard function at the posterior mean of $\beta$.
Solid lines give 25 posterior draws.
In both cases, the hazard functions have a smoothly decreasing shape with high posterior probability.
Although posterior precision is fairly high, there is still a non-negligible posterior probability that the hazard function is non-monotonic, at least for the POT series.
The smooth variation in the hazard function, together with the possibility of it being non-monotonic, would have been impossible to capture with commonly used parametric conditional duration densities.

\section{Conclusion}\label{sec:conclusion}
Models in the literature are designed to capture regular durations, those between unrelated trades.
They are not intended, nor well suited, to capture the observed clustering of related trades.
Common practice is to aggregate seemingly related trades into clusters and model only the ``regular'' durations between clusters.
Even if trades could be classified as related or not without error, it is not clear that this would be desirable since it involves discarding information relevant to liquidity measurement and market microstructure.
Furthermore, since it is not easy to tell related trades from unrelated trades that just happen to occur within the same second, errors of classification are inevitable.
The most common aggregation rule, the same-second rule, amounts to calling all durations recorded as $0s$ cluster durations, and all others, regular durations.
It is clear, however, that many of the durations recorded as $0s$ must be regular by happenstance---we just don't know which ones---and they are erroneously classified as cluster durations by the same-second rule.
One consequence is to understate trade intensity and liquidity, especially at times of high intensity.
Another is that the abrupt change in the number of durations between $0s$ and $1s$ makes it difficult to fit a normalized conditional duration distribution.
The rule suggested by \cite{GramWell02} clearly mitigates the problem, but it does not eliminate it.

The solution we proposed is to make our model a mixture model, with a binary state variable indicating which durations are cluster durations and which are regular.
Identification of the two states comes from the very tight distribution of cluster durations and, more subtly, the shrinkage of the normalized conditional duration density towards an exponential density, which varies slowly near zero.
This probabilistic, rather than deterministic, classification, allows us to learn that any given pair of trades recorded in the same second are very probably related, at the same time as we learn that many such pairs are not in fact pairs of related trades.

Despite not learning which durations recorded as $0s$ are regular, we are able to estimate quite precisely the normalized conditional duration density.
Its hazard function does not exhibit the large changes near zero that occur using the \cite{GramWell02} rule, which we claim is an artifact of the misclassification of many unrelated but nearly simultaneous trades as being related.

We introduced a flexible distribution for regular durations.
Appealing to queueing theory, we argued that a good first-order model for durations between unrelated trades is an exponential distribution.
We introduce a normalized conditional distribution for regular durations that is flexible, and also expressible as a perturbation of an exponential distribution.
This allows us to shrink towards the exponential distribution.

Due in part to efficient draws of the latent trade intensity state, and despite the flexible distribution, numerical efficiency of posterior simulation is considerably better than that of previous studies where duration distributions are parametric.

In the empirical application, we found that the conditional hazard function for regular durations varies much less than what is found in many studies.
We attribute this to better, probabilistic, classification of trades as related or not, and using flexible duration distributions instead of parametric distributions whose hazard functions have implausible behaviour near zero.

\bibliographystyle{apacite}
\bibliography{bibliography}

\appendix
\section{Derivatives}\label{app:derivatives}
We show here how to evaluate the first five derivatives of $\psi(\xdi) \equiv \log p(\ydi \vbar \sdi, \xdi)$ with respect to $\xdi$, at an arbitrary value of $\xdi$.
Routines for these derivatives are required by the HESSIAN method.
To avoid tedium and error, we do not provide analytic expressions for the derivatives.
Instead, we give derivatives of primitive functions and show how to combine them using F\'aa di Bruno's formula, a generalization of the chain rule to higher derivatives, to compute exact derivatives of $\psi(\xdi)$.
It gives derivatives as
\[
	\frac{d^n}{dx^n} f(g(x)) 
		= \sum \frac{n!}{m_1!m_2! \cdots m_n!} 
			f^{(m_1+\cdots+m_n)}(g(x))
			\prod_{j=1}^n \left(\frac{g^{(j)}(x)}{j!}\right)^{m_j}.
\]

For $\sdi = 0$, all derivatives of $\psi(\xdi)$ are equal to zero.
For $\sdi=1$, we use the representation of $p(\ydi \vbar \sdi, \xdi)$ as a linear combination of exponential densities in~\eqref{eq:linear-exp}.
This gives
\begin{equation}
	\psi(\xdi)
	= \log \left[
		\sum_{j=1}^J \alpha_j j \tilde \lambda e^{-\xdi-j \tilde \lambda e^{-\xdi}\ydi} \right] 
    = \log \left[ \sum_{j=1}^J \alpha_j \lambda_j g_j(\xdi) \right],
\end{equation}
where $\lambda_j = j \tilde \lambda$, $g_j(\xdi) = e^{h_j(\xdi)}$  and $h_j(\xdi) = -\xdi - \lambda_j e^{-\xdi}\ydi$ for $j=1,\ldots,J$.
We compute derivatives of $\psi(\xdi)$ bottom up. The steps are
\begin{enumerate}
    \item Compute $h_j(\xdi)$ and its first five derivatives with respect to $\xdi$, $j=1,\ldots,J$:
    \begin{align*}
    	h_j'(\xdi)   &= -1 + \lambda_j e^{-\xdi}\ydi,
    	&h_j'''(\xdi) &= h_j^{(5)}(\xdi) = \lambda_j e^{-\xdi}\ydi, \\
        h_j''(\xdi)  &= h_j^{(4)}(\xdi) = -\lambda_j e^{-\xdi}\ydi.
    \end{align*}

    \item Compute $g_j(\xdi)$ and first five derivatives with respect to $\xdi$, using F\'aa di Bruno's rule, for the exponential function composed with $h_j(\xdi)$.

    \item Compute $p(\ydi \vbar \sdi, \xdi)$ and first five derivatives with respect to $\xdi$, using
    \[
        \frac{\partial^{(r)}}{\partial \xdi^{(r)}}p(\ydi \vbar \sdi, \xdi) = 
        \sum_{j=1}^J \alpha_j \lambda_j g_j^{(r)}(\xdi),
        \qquad r=1,\ldots,5.
    \]
    \item Compute $\psi(\xdi)$ and first five derivatives with respect to $\xdi$, using
    F\'aa di Bruno's rule, for the logarithmic function composed with $p(\ydi \vbar \sdi, \xdi)$,
    treated as a function of $\xdi$.
    The first five derivatives of $q(z) = \ln z$ are
    \[
        q'(z) = z^{-1}, \quad q''(z) = -z^{-2}, \quad
        q'''(z) = 2z^{-3}, \quad q^{(4)}(z) = -6z^{-4}, \quad q^{(5)} = 24z^{-5}(z).
    \]
\end{enumerate}

\section{Drawing artificial observations}\label{app:drawing-y}
Here we show how to draw observations from their conditional distribution $p(y \vbar \beta, \lambda, \pi, \regime, \state)$.
For the latent state process used for the Getting it right experiment (see equation~\eqref{eq:dynamic-gir}), there is no relationship between the duration and the dynamic of the latent state.
In this case, updating the vector of observations can be done efficiently with the following Metropolis-Hastings algorithm:
\begin{itemize}
	\item For $i=1,\dots,n$, draw a proposal $\obs_i ^* \sim p(\obs_i \vbar \beta, \lambda, \pi, \regime_i, \state_i)$.
	\item Define $y_0^*=0$ and construct the corresponding transaction times $t_i^* = \sum_{k=0}^{i-1} \obs_k^*$ to evaluate the B-spline function $\mean(t)$.
	\item Accept the proposal $(\obs_1^*,\ldots,\obs_n^*)$ with probability
	\[
	\min
	\left\{
	1, \,
	\prod_{i=2}^n
	\frac
	{p\big(\state_i \vbar \phi, \sigma, \delta, \state_{i-1}, t_{i-1}, t_{i-2}\big)}
	{p\big(\state_i \vbar \phi, \sigma, \delta, \state_{i-1}, t_{i-1}^*, t_{i-2}^*\big)}
	\right\}.
	\]
\end{itemize}

\end{document}

%% file: notation-fscd.tex
\newcommand{\topen}{t_\mathrm{open}}
\newcommand{\tclose}{t_\mathrm{close}}
\newcommand{\tdf}{t_{d,0}}

\newcommand{\tdi}{t_{d,i}}
\newcommand{\tdim}{t_{d,i-1}}

\newcommand{\obs}{y}

\newcommand{\ydi}{\obs_{d,i}}

\newcommand{\state}{x}
\newcommand{\xdf}{\state_{d,1}}

\newcommand{\xdi}{\state_{d,i}}
\newcommand{\xdim}{\state_{d,i-1}}
\newcommand{\xdip}{\state_{d,i+1}}

\newcommand{\mean}{m}

\newcommand{\regime}{s}
\newcommand{\sdf}{\regime_{d,1}}

\newcommand{\sdi}{\regime_{d,i}}
\newcommand{\sdim}{\regime_{d,i-1}}
\newcommand{\sdip}{\regime_{d,i+1}}


%% file: descriptive-statistics.tex
RY & 40999 & 2.85 & 7.37 & 176 & 2.58 & 68.2 & 5.7 & 3.3 & 2.8 & 2.4 & 2.1 \\
POT & 31123 & 3.76 & 11.34 & 289 & 3.02 & 73.2 & 4.0 & 2.4 & 1.8 & 1.6 & 1.3 \\

%% file: results-getting-it-right.tex
     $\theta_1$ &  4.74e-05 & 1.06e-04 &  0.446 &  3.06e-05 & 1.06e-04 &  0.288  \\ 
     $\theta_2$ & -1.91e-04 & 1.10e-04 & -1.735 & -1.53e-04 & 1.06e-04 & -1.433  \\ 
         $\tau$ &  8.50e-03 & 8.99e-03 &  0.946 & -1.09e-02 & 9.08e-03 & -1.204  \\ 
     $v'\delta$ &  4.99e-06 & 4.99e-05 &  0.100 & -6.04e-05 & 5.04e-05 & -1.199  \\ 
      $\beta_1$ & -4.94e-05 & 2.39e-05 & -2.065 & -2.42e-05 & 2.45e-05 & -0.988  \\ 
      $\beta_2$ &  3.64e-05 & 2.23e-05 &  1.636 &  4.47e-06 & 2.26e-05 &  0.198  \\ 
      $\beta_3$ &  1.30e-05 & 1.88e-05 &  0.692 &  1.97e-05 & 1.90e-05 &  1.035  \\ 
     $\xi_{00}$ &  3.81e-05 & 2.79e-05 &  1.364 &  1.25e-05 & 2.36e-05 &  0.530  \\ 
     $\xi_{11}$ &  1.56e-05 & 3.00e-05 &  0.522 & -1.57e-05 & 2.13e-05 & -0.739  \\ 
    $\lambda_1$ &  1.07e-03 & 4.52e-03 &  0.236 & & & \\ 
    $\lambda_2$ &  8.08e-05 & 2.25e-03 &  0.036 & & & \\ 
          $\pi$ & -2.89e-05 & 2.28e-05 & -1.268 & & & \\ 
        $\zeta$ &  &  &  & -1.01e-05 & 1.00e-05 & -1.003  \\ 

%% file: results-param-all-ry.tex
    $\phi^{\dagger}$ &  0.296 & 5.83e-02 & 0.135 &  0.293 & 5.84e-02 & 0.274 &  0.294 & 5.88e-02 & 0.318 &  0.295 & 5.91e-02 & 0.299 \\
            $\sigma$ &  0.399 & 1.92e-02 & 0.532 &  0.399 & 1.95e-02 & 0.435 &  0.399 & 1.95e-02 & 0.427 &  0.399 & 1.95e-02 & 0.417 \\
                     &  &  &  &  &  &  &  &  &  &  &  &  \\
           $\beta_1$ &  0.632 & 2.53e-02 & 0.248 &  0.428 & 1.67e-02 & 0.113 &  0.316 & 2.38e-02 & 0.060 &  0.265 & 2.14e-02 & 0.038 \\
           $\beta_2$ &  0.368 & 2.53e-02 & 0.248 &  0.301 & 2.34e-02 & 0.087 &  0.263 & 2.93e-02 & 0.063 &  0.173 & 3.44e-02 & 0.060 \\
           $\beta_3$ &  &  &  &  0.272 & 2.46e-02 & 0.220 &  0.212 & 2.45e-02 & 0.151 &  0.243 & 3.37e-02 & 0.128 \\
           $\beta_4$ &  &  &  &  &  &  &  0.208 & 2.48e-02 & 0.210 &  0.134 & 2.45e-02 & 0.131 \\
           $\beta_5$ &  &  &  &  &  &  &  &  &  &  0.186 & 2.26e-02 & 0.146 \\
                     &  &  &  &  &  &  &  &  &  &  &  &  \\
          $\xi_{00}$ &  0.743 & 3.16e-03 & 0.284 &  0.742 & 3.32e-03 & 0.144 &  0.742 & 3.44e-03 & 0.100 &  0.742 & 3.51e-03 & 0.079 \\
          $\xi_{11}$ &  0.498 & 5.04e-03 & 0.547 &  0.500 & 5.23e-03 & 0.294 &  0.500 & 5.34e-03 & 0.197 &  0.501 & 5.44e-03 & 0.119 \\
             $\zeta$ &  0.985 & 1.76e-03 & 0.085 &  0.986 & 2.03e-03 & 0.048 &  0.986 & 2.21e-03 & 0.039 &  0.987 & 2.28e-03 & 0.030 \\

%% file: results-param-reg-ry.tex
    $\phi^{\dagger}$ &  0.333 & 7.07e-02 & 0.367 &  0.364 & 7.77e-02 & 0.433 &  0.375 & 8.05e-02 & 0.432 &  0.383 & 8.21e-02 & 0.367 \\
            $\sigma$ &  0.375 & 1.81e-02 & 0.484 &  0.378 & 1.77e-02 & 0.569 &  0.374 & 1.73e-02 & 0.551 &  0.372 & 1.72e-02 & 0.570 \\
                     &  &  &  &  &  &  &  &  &  &  &  &  \\
           $\beta_1$ &  0.570 & 2.37e-02 & 0.372 &  0.379 & 1.81e-02 & 0.417 &  0.195 & 1.57e-02 & 0.394 &  0.135 & 1.46e-02 & 0.352 \\
           $\beta_2$ &  0.430 & 2.37e-02 & 0.372 &  0.397 & 1.68e-02 & 0.504 &  0.424 & 2.18e-02 & 0.557 &  0.364 & 2.73e-02 & 0.356 \\
           $\beta_3$ &  &  &  &  0.225 & 2.38e-02 & 0.661 &  0.115 & 2.19e-02 & 0.510 &  0.139 & 3.04e-02 & 0.391 \\
           $\beta_4$ &  &  &  &  &  &  &  0.266 & 2.09e-02 & 0.437 &  0.160 & 2.38e-02 & 0.437 \\
           $\beta_5$ &  &  &  &  &  &  &  &  &  &  0.202 & 2.22e-02 & 0.429 \\

%% file: results-param-all-pot.tex
    $\phi^{\dagger}$ &  0.466 & 9.96e-02 & 0.344 &  0.396 & 8.67e-02 & 0.287 &  0.400 & 8.77e-02 & 0.275 &  0.404 & 8.85e-02 & 0.253 \\
            $\sigma$ &  0.453 & 2.11e-02 & 0.384 &  0.442 & 2.16e-02 & 0.391 &  0.442 & 2.18e-02 & 0.387 &  0.442 & 2.18e-02 & 0.356 \\
                     &  &  &  &  &  &  &  &  &  &  &  &  \\
           $\beta_1$ &  0.708 & 3.17e-02 & 0.466 &  0.502 & 2.21e-02 & 0.252 &  0.362 & 2.96e-02 & 0.105 &  0.286 & 2.72e-02 & 0.084 \\
           $\beta_2$ &  0.292 & 3.17e-02 & 0.466 &  0.252 & 2.33e-02 & 0.165 &  0.268 & 3.36e-02 & 0.103 &  0.227 & 3.62e-02 & 0.092 \\
           $\beta_3$ &  &  &  &  0.246 & 2.67e-02 & 0.268 &  0.171 & 2.85e-02 & 0.179 &  0.188 & 3.53e-02 & 0.147 \\
           $\beta_4$ &  &  &  &  &  &  &  0.199 & 2.84e-02 & 0.213 &  0.131 & 2.59e-02 & 0.162 \\
           $\beta_5$ &  &  &  &  &  &  &  &  &  &  0.169 & 2.67e-02 & 0.165 \\
                     &  &  &  &  &  &  &  &  &  &  &  &  \\
          $\xi_{00}$ &  0.788 & 3.03e-03 & 0.505 &  0.786 & 3.14e-03 & 0.355 &  0.787 & 3.18e-03 & 0.247 &  0.787 & 3.24e-03 & 0.202 \\
          $\xi_{11}$ &  0.441 & 6.08e-03 & 0.612 &  0.444 & 6.16e-03 & 0.577 &  0.443 & 6.22e-03 & 0.408 &  0.443 & 6.25e-03 & 0.342 \\
             $\zeta$ &  0.985 & 1.46e-03 & 0.170 &  0.986 & 1.57e-03 & 0.107 &  0.986 & 1.67e-03 & 0.088 &  0.986 & 1.75e-03 & 0.071 \\

%% file: results-param-reg-pot.tex
    $\phi^{\dagger}$ &  0.483 & 1.02e-01 & 0.586 &  0.385 & 8.31e-02 & 0.414 &  0.412 & 8.95e-02 & 0.398 &  0.419 & 9.15e-02 & 0.348 \\
            $\sigma$ &  0.434 & 2.02e-02 & 0.684 &  0.419 & 2.11e-02 & 0.507 &  0.419 & 2.07e-02 & 0.498 &  0.419 & 2.06e-02 & 0.465 \\
                     &  &  &  &  &  &  &  &  &  &  &  &  \\
           $\beta_1$ &  0.718 & 2.99e-02 & 0.644 &  0.507 & 2.33e-02 & 0.409 &  0.329 & 2.39e-02 & 0.341 &  0.247 & 2.09e-02 & 0.313 \\
           $\beta_2$ &  0.282 & 2.99e-02 & 0.644 &  0.271 & 1.91e-02 & 0.442 &  0.324 & 2.69e-02 & 0.446 &  0.296 & 2.87e-02 & 0.358 \\
           $\beta_3$ &  &  &  &  0.222 & 2.49e-02 & 0.657 &  0.130 & 2.60e-02 & 0.447 &  0.146 & 3.19e-02 & 0.372 \\
           $\beta_4$ &  &  &  &  &  &  &  0.217 & 2.58e-02 & 0.398 &  0.132 & 2.49e-02 & 0.381 \\
           $\beta_5$ &  &  &  &  &  &  &  &  &  &  0.178 & 2.51e-02 & 0.339 \\

%% file: results-half-life.tex
  FSCD(2) & 243.8 & 49.2 & 163.8 & 354.7 & 155.5 & 33.7 & 100.8 & 232.0 \\
  FSCD(3) & 245.8 & 49.1 & 164.0 & 355.2 & 183.7 & 40.5 & 116.9 & 274.3 \\
  FSCD(4) & 245.0 & 49.2 & 163.8 & 355.3 & 181.8 & 40.1 & 116.6 & 273.6 \\
  FSCD(5) & 244.6 & 49.4 & 163.1 & 355.1 & 179.7 & 39.4 & 115.1 & 267.9 \\
          &  &  &  &  &  &  &  &  \\
R-FSCD(2) & 217.4 & 46.1 & 141.1 & 321.0 & 149.9 & 31.9 & 97.5 & 222.1 \\
R-FSCD(3) & 199.1 & 42.5 & 128.6 & 295.8 & 188.5 & 40.8 & 121.2 & 281.9 \\
R-FSCD(4) & 193.4 & 41.4 & 125.0 & 285.4 & 176.4 & 38.6 & 113.6 & 262.9 \\
R-FSCD(5) & 189.3 & 40.5 & 122.2 & 278.4 & 173.1 & 37.7 & 110.9 & 257.0 \\

%% file: results-classification.tex
     $0s$ &   FSCD(2) & 1118 &   40 & 1040 & 1198 &  501 &   25 &  452 &  552 \\
          &   FSCD(3) & 1156 &   50 & 1060 & 1254 &  543 &   29 &  487 &  601 \\
          &   FSCD(4) & 1150 &   57 & 1040 & 1264 &  536 &   32 &  473 &  601 \\
          &   FSCD(5) & 1180 &   63 & 1058 & 1304 &  536 &   35 &  468 &  607 \\
          &           &  &  &  &  &  &  &  &  \\
     $1s$ &   FSCD(2) & 1936 &   44 & 1851 & 2021 &  905 &   27 &  851 &  959 \\
          &   FSCD(3) & 1972 &   52 & 1869 & 2071 &  949 &   31 &  888 & 1010 \\
          &   FSCD(4) & 1967 &   57 & 1856 & 2080 &  943 &   34 &  876 & 1009 \\
          &   FSCD(5) & 1987 &   59 & 1873 & 2104 &  943 &   36 &  874 & 1013 \\